\documentclass[runningheads]{llncs}
\usepackage{graphicx}
\usepackage{caption}

\usepackage{amsmath}
\usepackage{amsmath,amssymb,amsfonts}
\usepackage{mathtools}
\usepackage{multirow}
\usepackage{setspace}
\begin{document}

	\title{Improving security and bandwidth efficiency of NewHope using error-correction schemes	\thanks{Supported by X.}}
	
	%
	%\titlerunning{Abbreviated paper title}
	% If the paper title is too long for the running head, you can set
	% an abbreviated paper title here
	%
	\author{Minki Song\inst{1}\and
		Seunghwan Lee\inst{1} \and
		Eunsang Lee\inst{2} \and
		Dong-Joon Shin\inst{1}	\and	
		Young-Sik Kim\inst{3} \and
		Jong-Seon No\inst{2}
		}
	\authorrunning{M. Song et al.}
	% First names are abbreviated in the running head.
	% If there are more than two authors, 'et al.' is used.
	%
	\institute{Hanyang University, Seoul, Korea \and
	Seoul National University, Seoul, Korea \and
	Chosun University, Gwang-ju, Korea \\
	%Springer Heidelberg, Tiergartenstr. 17, 69121 Heidelberg, Germany
	\email{minkisong@hanyang.ac.kr, kr3951@hanyang.ac.kr, eslee3209@ccl.snu.ac.kr, djshin@hanyang.ac.kr, iamyskim@chosun.ac.kr, jsno@snu.ac.kr}}
	%\url{http://www.springer.com/gp/computer-science/lncs} \and
	%ABC Institute, Rupert-Karls-University Heidelberg, Heidelberg, Germany\\
	%\email{\{abc,lncs\}@uni-heidelberg.de}}
	
\maketitle              % typeset the header of the contribution

\begin{abstract}
Among many submissions to the NIST post-quantum cryptography (PQC) project, NewHope is a promising key encapsulation mechanism (KEM) based on the Ring-Learning with errors (Ring-LWE) problem.
Since the most important factors to be considered for PQC are security and cost including bandwidth and time/space complexity, in this paper, by doing exact noise analysis and using Bose Chaudhuri Hocquenghem (BCH) codes, it is shown that the security and bandwidth efficiency of NewHope can be substantially improved.
In detail, the decryption failure rate (DFR) of NewHope is recalculated by performing exact noise analysis, and it is shown that the DFR of NewHope has been too conservatively calculated.
Since the recalculated DFR is much lower than the required $2^{-128}$, this DFR margin is exploited to improve the security up to 8.5 \% or the bandwidth efficiency up to 5.9 \% without changing the procedure of NewHope.

The additive threshold encoding (ATE) used in NewHope is a simple error correcting code (ECC) robust to side channel attack, but its error-correction capability is relatively weak compared with other ECCs.
Therefore, if a proper error-correction scheme is applied to NewHope, either security or bandwidth efficiency or both can be improved.
Among various ECCs, BCH code has been widely studied for its application to cryptosystems due to its advantages such as no error floor problem.
In this paper, the ATE and total noise channel are regarded as a super channel from an information-theoretic viewpoint.
Based on this super channel analysis, various concatenated coding schemes of ATE and BCH code for NewHope have been investigated.
Through numerical analysis, it is revealed that the security and bandwidth efficiency of NewHope are substantially improved by using the proposed error-correction schemes. 

%To this end, the security and bandwidth efficiency of NewHope can be improved by concatenating Bose–Chaudhuri–Hocquenghem (BCH) code having strong error correction capability and ATE.
%In results, Applying BCH code to NewHope improves security by up to 21.5 \%, or by reducing ciphertext by up to 35.3 \%.

%PQC에서 가장 중요한 요소는 security이고, 그 다음으로 중요한 것은 cost와 performance이다. 이러한 측면에서 NIST에 제출된 NewHope는 훌륭한 PQC 후보이다. 그러나 NewHope에서는 DFR 계산이 정확하게 이루어지지 않았다. 본 논문에서는 정확한 noise 분석을 통해 NewHope의 DFR을 재계산하고 이를 이용하여 복잡도 증가 없이 security를 8.5 \% 개선시키거나 ciphertext size를 5.9 \% 감소시켜 bandwidth efficiency를 개선시킨다. 추가적으로 약간의 복잡도를 증가시켜 ciphertext size를 23.5 \% 감소시켜 bandwidth efficiency를 상당히 개선시켰다. 여기에 더 나아가 강력한 오류 정정 부호인 BCH code를 적용시켜 security를 최대 21.5 \% 개선시키고, 또는 ciphertext를 최대 35.3 \% 감소시켜 bandwidth efficiency를 개선시켰다.

	%Proceeding..	
	%(The abstract should briefly summarize the contents of the paper in
	%150--250 words.)

	\keywords{Bandwidth Efficiency \and BCH Code \and Decryption Failure Rate \and Error Correcting Codes \and NewHope \and NIST \and Post-Quantum Cryptography \and Security}
\end{abstract}

\section{Introduction}

%We define \xi_k to be the rounded Gaussian distribution of parameter $\sigma=\sqrt{k/2}$, that is the distribution of $\lfloor \sqrt{k/2} \cdot x \rceil$ where $x$ follows the standard normal distribution.
% 그림은 k가 증가할수록 \psi_k와 \xi_k의 distance가 감소함을 보여준다. 비록 k의 증가는 NewHope의 time complexity를 증가시키지만, ECC 적용으로 k를 증가시킨다면 NewHope의 security를 개선시킬 수 있다.

Current public-key algorithms based on integer decomposition, discrete logarithm, and elliptic curve discrete logarithm problems (e.g, RSA and elliptic curve encryption) have been unlikely to be broken by currently available technology.
However, with the help of upcoming quantum computing technology such as Shor's quantum algorithm for integer factorization, current public-key algorithms can be easily broken.
For that reason, in order to avoid the security problem of future network, new public-key algorithms called post-quantum cryptography (PQC) should be developed to replace the existing public-key algorithms.
Therefore, the National Institute of Standards and Technology (NIST) has recently begun a PQC project to identify and evaluate post-quantum public-key algorithms secure against quantum computing \cite{NEON1}.
Among the various PQC candidates, lattice-based cryptosystems have become one of the most promising candidate algorithms for post-quantum key exchange.
Lattice-based cryptosystems have been developed based on worst-case assumptions about lattice problems that are believed to be resistant to quantum computing.
Among various lattice problems, learning with errors (LWE) problem introduced by Regev in 2005 \cite{LWE} has been widely analyzed and used.
Furthermore, the Ring-LWE problem presented by Lynbashevsky, Peikert, and Regev in 2010 \cite{RLWE}, which improves the computational and implementation efficiency of LWE, has also been widely used \cite{NewHope NIST}, \cite{Frodo}, \cite{Kyber}, \cite{LAC}, \cite{HILA5}. 
NewHope has been proposed by Alkim, Ducas, P{\"o}ppelmann, Schwabe \cite{NewHope1}, \cite{NewHope2} which is one of the various cryptosystems based on Ring-LWE.
NewHope has attracted a lot of attention \cite{NewHope Gaborit}, \cite{NEON}, \cite{Fritzmann} and it was verified in an experiment of Google \cite{NEON 8}.
The key reasons that NewHope attracts so much attention are the use of simple and practical noise distribution, a centered binomial distribution, and a proper choice of ring parameters for better performance and security.

% NIST에 제안된 NewHope는 [NewHope Simple]의 약간 조정된 version이다. (Noise variance parameter 변화, 복잡도를 낮추기 위한 NTT 사용 등) 

NewHope is an indistinguishability (IND)-chosen ciphertext attack (CCA) secure key encapsulation mechanism (KEM) that exchanges the shared secret key based on the IND-chosen plaintext attack (CPA) secure public-key encryption (PKE).
Note that the IND-CPA PKE can be transformed into the IND-CCA KEM using Fujisaki-Okamoto (FO) transform \cite{FO}.
The IND-CCA secure KEM obtained by applying FO transform to IND-CPA secure PKE requires a very low decryption failure rate (DFR) because an attacker can exploit the decryption error \cite{FO}.
Therefore, the DFR in NewHope should be lower than $2^{-128}$ to make sure of resilience against attacks that exploit decryption errors.
As in Frodo \cite{Frodo} and Kyber \cite{Kyber}, this study aims to achieve the DFR lower than $2^{-140}$ to allow enough margin in NewHope.
%NewHope는 IND-CPA PKE를 기반으로 shared secret key를 교환하는 IND-CCA KEM 이다. 여기서 널리 사용되고 있는 Fujisaki-Okamoto transformation를 활용하여 IND-CPA PKE를 IND-CCA KEM으로 변환한다. 

%DFR는 noise에 영향을 받으며, 거기에 포함된 error를 정정하는 ECC와 밀접한 연관이 있다. 
The DFR of NewHope is most influenced by the noise parameter $k$ of centered binomial distribution and modulus $q$, and is also closely related to an error correcting code (ECC).
NewHope uses an additive threshold encoding (ATE) as an ECC, which is almost similar to the repetition code used in digital communication systems \cite{ATE}.
In addition, ATE has an advantage of being robust to the side channel attack, but ATE inherently has much worse error-correction capability than other ECCs such as Bose Chaudhuri Hocquenghem (BCH) code, low-density parity-check (LDPC) codes, and turbo codes \cite{Moon}, \cite{Wicker}.
In \cite{Fritzmann}, it was shown that by applying more powerful ECC to NewHope instead of ATE, the DFR can be improved. 
It is also shown that the security can be improved and the size of ciphertext can be reduced by using the improved DFR obtained by using ECC.
%NewHope에는 더 좋은 security와 decryption failure rate (DFR)을 위해서 error correcting code (ECC) 역할을 하는 additive threshold encoding (ATE)가 적용되어 있다. ATE는 encoding과 decoding이 아주 간단한 repetition code이다. 단, 기존 repetition code의 복호와 달리 반복된 값들에 절댓값을 취한 후 더해주는 방식으로 복호가 이루어진다. ATE의 구현이 아주 간단하면 side channel attack에 강인한 장점이 있다. 그러나 낮은 오류 정정 능력을 갖는 단점이 있다. 따라서 더 강력한 ECC를 적용시킨다면 security와 DFR을 개선시킬 수 있다. 추가적으로, 개선된 DFR을 활용하여 ciphertext size를 감소시킬 수 있으며 이를 통해 bandwidth efficiency를 개선시킬 수 있다.
%NewHope에는 error correcting code (ECC)로써 additive threshold encoding (ATE)가 적용되어 있다. 

\subsubsection{Contribution}
The contributions of this paper is divided into six categories.

-\textbf{Exact Noise Analysis of NewHope}
NewHope can be understood as a digital communication system.
Bob and Alice are transmitter and receiver, respectively, and the 256-bit shared secret key is a message bit stream. 
The difference between the ATE output $v$ and the received signal $v''$ distorted by many factors can be modeled as a digital communication channel.
We analyze all the noise sources of this channel and numerically calculate the exact noise distribution of NewHope.
%우리는 이 channel을 수학적을 분석하여 noise를 difference noise, compression noise(ciphertext $v'$)  

%Difference noise, compression noise, total noise

-\textbf{Recalculation of DFR of NewHope}
The DFR of NewHope is recalculated as $2^{-474}$ and $2^{-431}$ for $n=1024$ and $n=512$, respectively, based pm the exact noise analysis and theoretical analysis of ATE.
New DFR values show that the DFR of NewHope has been too conservatively calculated.

-\textbf{Improvement of Security and Bandwidth Efficiency of NewHope Using New DFR Margin}
Since the recalculated DFR is much lower than the required $2^{-140}$, this DFR margin is exploited to improve the security by 8.5 \% or bandwidth efficiency by 5.9 \% without changing the procedure of NewHope.
If a slight increase in time/space complexity is allowed, the bandwidth efficiency can be improved by 23.5 \%.

-\textbf{An Information-theoretic View of Noise in NewHope as a Super Channel}
The ATE and total noise of NewHope can be regarded as a super channel from an information-theoretic viewpoint.
For various concatenated coding schemes of ATE and BCH code for NewHope, super channel is defined.
We perform exact analysis of super channel for each concatenated coding schemes by using exact analysis of noise and ATE.

-\textbf{Proposed Concatenated Coding Schemes}
% BCH code 선정 이유 (LDPC, Turbo code 제외한 이유), ATE와 연접한 이유
% n=1024에 3가지, n=512에 한가지 제안
The ATE used in NewHope is simple and robust to the side channel attack.
However, since ATE is based on the repetition code, it shows low error-correction capability. 
In order to improve the security and bandwidth efficiency, advanced ECCs should be applied to NewHope.
BCH codes are suitable for PQC because they do not show an error floor problem and can be analyzed by deriving an upper bound of block error rate (BLER).
Based on extensive simulations, we select four concatenated coding schemes of ATE and BCH code to combine the advantages of these two ECCs.

-\textbf{Analysis of Security and Bandwidth Efficiency of NewHope with the Proposed Error-correction Schemes}
The results of numerical analysis show that the proposed concatenated coding schemes of ATE and BCH code can improve the security level by 21.5 \% or reduce the ciphertext size by 41.5 \% while achieving the required DFR $2^{-140}$ for NewHope with $n=1024$.
Also, for NewHope with $n=512$, the proposed concatenated coding schemes can improve the security level by 22.8 \% or reduce the ciphertext size by 35.3 \% while achieving the required DFR $2^{-140}$.

The contributions of this paper differ from the contributions of \cite{Fritzmann} in three ways.
First, we interpret NewHope as a communication system over noisy channel from an information-theoretic view, and hence a super channel is defined and analyzed in performing exact analysis of noise and ATE.
Based on this analysis, we recalculate the exact DFR of NewHope.
Second, we show that by using the recalculated DFR, the security and bandwidth efficiency of NewHope can be improved without changing the procedure of NewHope. 
Finally, compared with \cite{Fritzmann}, we numerically evaluate more various compression rates of ciphertext using the super channel analysis. 
Through this evaluation, we present many effective compression rates show that better performance in reducing the size of the ciphertext and decreasing show lower DFR.
 
% 본 논문에서는 NewHope의 security와 bandwidth efficiency를 개
%The most important factor of PQC is security, followed by cost and performance.
%The cost and performance are the size of public key, secret key, and ciphertext and DFR.
%We define a new parameter that is a compression rate.
%The compression rate as the new parameter of NewHope can be exploited to expand the concept of security and bandwidth efficiency, and through various calculations and simulations the compression rate is suitably determined as PQC.
%NewHope submitted to NIST has a very conservative calculation of DFR calculations, which use Chernoff-Cramer bound and an approximation and 
%In this paper, we recalculate DFR through an exact noise analysis of NewHope, and propose to increase security level or improve bandwidth efficiency without increasing complexity.
%Also, we propose to improve security level and bandwidth efficiency by increasing some complexity.
%A study is conducted to improve security and bandwidth efficiency by concatenating ATE applied as ECC with NewHope with BCH code with strong error correction capability.

%PQC의 가장 중요한 요소는 security이고 그 다음으로는 cost와 performance이다. cost와 performance로는 the size of pulic key, secret key, and ciphertext and DFR. NIST에 제출된 NewHope은 DFR 계산이 아주 보수적으로 계산되어 있다.  본 논문에서는 NewHope의 정확한 noise 분석을 통해 과도하게 책정된 DFR을 파악하고, 이를 활용하여 복잡도 증가가 거의 없이 security level을 증가시키거나 bandwidth efficiency를 개선시키는 방법을 제안한다. 
%NewHope에 ECC로써 적용된 ATE를 강력한 오류정정능력을 갖는 BCH code와 연접하여 security와 bandwidth efficiency를 개선시키는 연구를 수행한다.

\section{Preliminaries}
\subsection{NewHope}
\subsubsection{Parameters}
There are three important parameters in NewHope: $n$, $q$, and $k$. 
The dimension $n=512$ or $1024$ for NewHope guarantees the security properties of Ring-LWE and enables efficient number theoretic transform (NTT) \cite{NTT}. 
The modulus $q=12289$ is determined to support security and efficient NTT, and is closely related with the bandwidth. 
The noise parameter $k=8$ is the parameter of centered binomial distribution, which determines the noise strength and hence directly affects the security and DFR \cite{NewHope NIST}.

%In this paper, a new parameter called the compression rate $r=(r_{v'},r_{\hat{u}})$ as new parameter is additionally define in NewHope.
%The compression rate is related to the ciphertext size and DFR, which can be exploited to expand the security and bandwidth efficiency of NewHope.
%In NewHope, the compression rate is set to $r=(r_{v'}=8,r_{\hat{u}}=q)$.
%NewHope의 parameter로는 n, q, k로 세가지가 있다.  n은 ~와 관련있고 512, 1024로 정해진다. q는 ~와 관련있고, ~때문에 12289로 정해진다. k는 centered binomial distribution의 parameter로써 noise variance를 결정하여 security level과 DFR에 영향을 준다. k가 클수록 security level이 증가하지만 DFR도 함께 증가하게 된다. 

\subsubsection{Notations}
Let $\mathcal{R}_q=\mathbb{Z}_q[x]/(X^n+1)$ be the ring of integer polynomials modulo $X^n+1$ where each coefficient is reduced modulo $q$.
Let $a \xleftarrow{\text{\$}} \chi $ be the sampling of $a \in \mathcal{R}_q$ following the probability  distribution $\chi$ over $\mathcal{R}_q$.
Let $\psi_{k}$ denote the centered binomial distribution with parameter $k$, which is practically realized by $\sum_{i=0}^{k-1}(b_i-b_i')$, where $b_i$ and $b_i'$ are uniformly and independently sampled from $\{0,1\}$. 
The variance of $\psi_{k}$ is $k/2$ \cite{NewHope NIST}.
$a \circ b$ denotes the coefficient-wise product of polynomials $a$ and $b$.

\subsubsection{NewHope Protocol} 
\begin{figure}
\centering
\includegraphics[width=\textwidth]{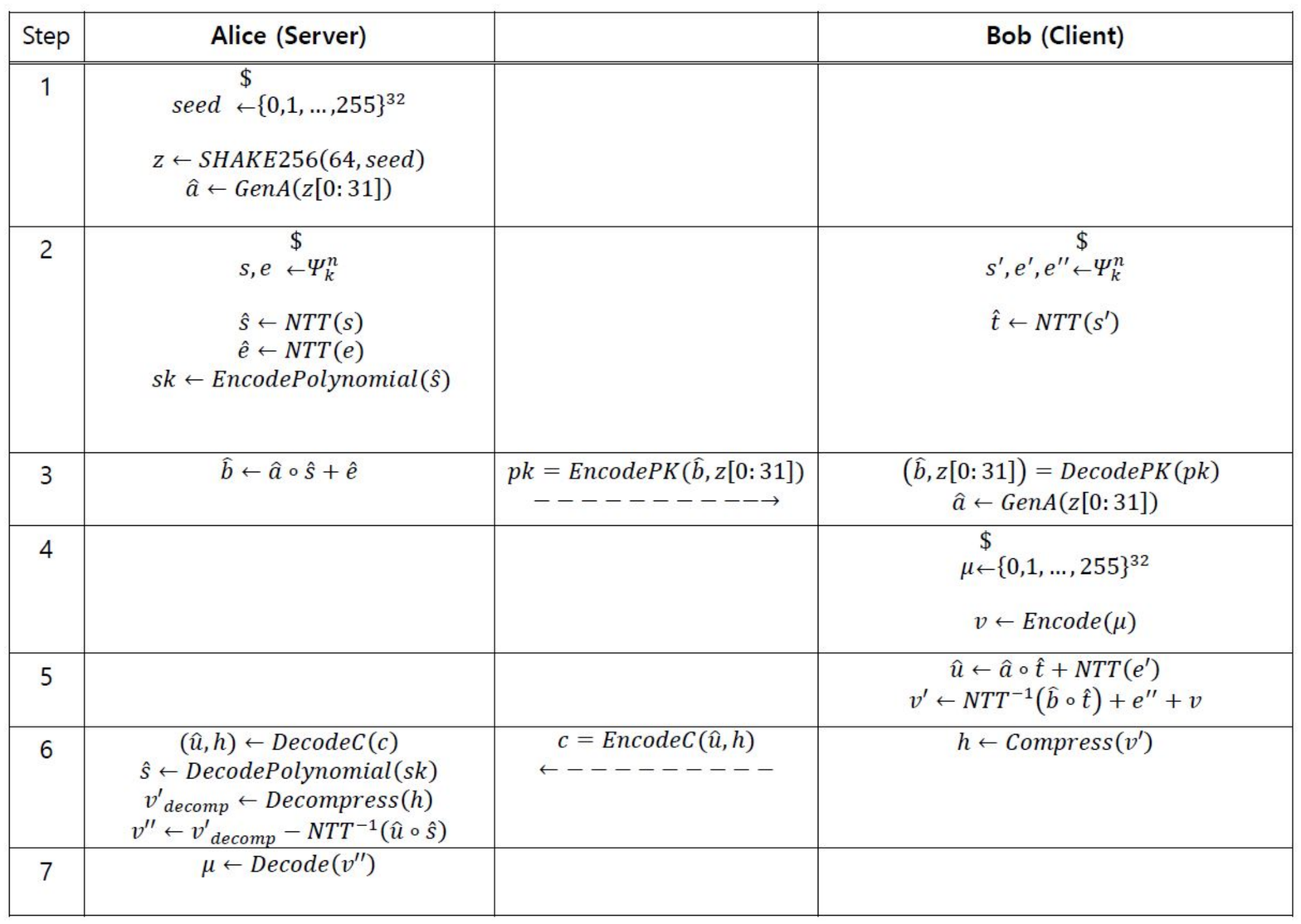}
\caption{NewHope Protocol.} 
\label{NewHopeProtocol}
\end{figure}

% NewHope는 Alice(Server)와 Bob(Client)가 서로 secret key를 공유하기 위한 Lattice 기반의 KEM cryptosystem이다. NewHope의 protocol은 다음과 같이 간단히 표현할 수 있다. 
NewHope is a lattice-based KEM for Alice (Server) and Bob (Client) to share 256-bit secret key with each other. 
The protocol of NewHope is briefly expressed based on Fig. \ref{NewHopeProtocol} as follows, where the functions are the same ones as defined in \cite{NewHope NIST}. \\

Step 1) $seed \xleftarrow{\text{\$}} \{0,1,\dots, 255\}^{32} $ denotes a uniformly sampling of 32 byte arrays (corresponding to 256 bits) with 32 integer elements selected between 0 and 255 by using a random number generator. 
Then $SHAKE256(l,d)$, a strong hash function \cite{SHAKE256}, takes an integer $l$ that specifies the number of output bytes and an data byte array $d$ as inputs. 
In NewHope, $z \leftarrow SHAKE256(64,seed)$ denotes  that 32 byte arrays $seed$ are hashed to generate a pseudorandom 64 byte arrays $z$ with 64 integer elements uniformly selected between 0 and 255.
Then $GenA$ expands the pseudorandom 32 byte arrays $z[0:31]$ using $SHAKE128$ hash function \cite{SHAKE256} to generate the polynomial $\hat{a} \in \mathcal{R}_q $ where $z[0:31]$ is the first 32 byte arrays of $z$.
Since $\hat{a}$ is generated from the $seed$ sampled from a uniform distribution, the coefficients of $\hat{a}$ also follow a uniform distribution on $[0,q-1]$.

Step 2) Generate polynomials ($s$, $s'$, $e$, $e'$, $e''$ $\in \mathcal{R}_q$) whose coefficients are sampled from the centered binomial distribution $\psi_k$. 
The polynomials ($s$, $s'$, $e$) are transformed to ($\hat{s}$, $\hat{t}$, $\hat{e}$), respectively, by applying NTT for efficient polynomial multiplication.
Then Alice transforms the secret key ($ \hat{s} $) into byte arrays using $ EncodePolynomial () $, which converts the polynomial ($ \hat{s} $) into 2048 byte arrays.
%곱해지는 Polynomial들은 효율적인 계산을 위해 NTT domain으로 mapping하고, 이 중에 Alice가 가지고 있을 secret key ($\hat{s}$)를 $EncodePolynomial()$을 이용하여 byte array로 변형한다.

Step 3) Alice creates a public key ($pk$) by converting $ \hat{b}  = \hat{a} \circ \hat{s} + \hat{e}$ and $z[0:31]$ into 1824 byte arrays using $ EncodePK () $, and transmits ($pk$) to Bob.
Then Bob transforms the received public key ($pk$) into ($ \hat{b} $, $z[0:31]$) using $ DecodePK () $, and creates ($ \hat{a} $), which is the same ($\hat{a}$) generated in Step 1.
%Alice는 public key ($\hat{b}$를 생성하고, $EncodePK()$를 이용하여 polynomial들을 byte array로 변환하여 Bob에게 전송한다. Bob은 수신된 public key를 $DecodePK()$로 polynomial 형태로 변형한 후 step 1의 Alice와 동일한 $\hat{a}$를 생성한다.

Step 4) A 256-bit shared secret key ($\mu$) is created by performing ATE encoding. 
%After mapping each bit of $\mu$ to the element in $\mathbb{Z}_q$ by using the rule $0 \in \mathbb{Z}_2 \rightarrow 0 \in \mathbb{Z}_q$ and $1 \in \mathbb{Z}_2 \rightarrow \lfloor \frac{q}{2} \rfloor \in \mathbb{Z}_q$, and it is repeatedly stored for multiple coefficients of polynomial ($v$).
%256-bit shared secret key($\mu$)를 생성하고 $Encode()$로 $0 \in \mathbb{Z}_2 \rightarrow 0 \in \mathbb{Z}_q$ , $1 \in \mathbb{Z}_2 \rightarrow \llcorner \frac{q}{2} \lrcorner \in \mathbb{Z}_q$

Step 5) Generate a ciphertext ($\hat{u}$, $v'$) using the public key and $v$.

Step 6) To efficiently reduce bandwidth, compression is performed on the coefficients of $ v '$ to generate the polynomial $h$, and then the ciphertext polynomials ($\hat{u}$, $h$) are transformed into the byte arrays $c$ using $EncodeC()$, and $c$ is transmitted to Alice.
Alice performs decompression on $\hat{h}$ to restore $v'$.
However, this restored polynomial $v'_{decomp}$ is different from $v'$ generated in Step 5, due to the compression and decompression.
Alice creates $ v '' $ using the received ciphertext $c$ and $ sk $ generated in Step 2. 
Each coefficient of $ v '' $ is a sum of the corresponding coefficient of $ v $ and noise. 
Note that $v'' $ is not a polynomial used in NewHope, for easy explanation of the results in this paper, $v''$ is added in Fig. \ref{NewHopeProtocol}.

Step 7) The 256-bit shared secret key ($ \mu $) is recovered (or decrypted) from the coefficients of $v''$ by performing ATE decoding.

%\subsubsection{Security level and bandwidth of NewHope}
%For security analysis of NewHope we mainly rely on the concrete security analysis of Ring-LWE based cryptosystems from [9]. Additionally, we also estimate the security level with the approach presented in [7]

% NewHope의 security 분석은 Ring-LWE based cryptosystems from [9]의 security analysis가 되었다. 그리고 security level을 추정하기 위해서 

%Primal attack and dual attack
% NewHope

%\subsubsection{Compression, ATE as ECC}
%ATE as ECC
%% ATE는 ECC 측면에서 볼때 굉장히 간단히 scheme이다. secret key의 하나의 bit를 4번 repetition하고, 0은 0으로 1은 q/2로 mapping하기만 한다. 그렇기 때문에 256 bits의 secret key를 ciphertext로 변환하는 과정을 통해 14*4*1024 bit가 필요하게 된다. 과하게 증가된 ciphertext size를 감소시키기 위해 low bit들에는 information이 거의 없다는 것을 활용하여 compression을 수행하여 ciphertext size를 감소시켜 Alice와 Bob이 서로 ciphertext를 교환한다. 
%
%Explanation about compression
%% Ciphertext에 대한 compression은 각 polynomial의 coefficient둘 별로 A를 통해 수행된다. 여기서 r은 compression rate를 나타내며, log2(r)로 coefficient의 크기가 감소된다. 이러한 과정을 통해서 비교적 효율적으로 ciphertext를 주고 받을 수 있다. 여기서 copmression rate에 대한 분석은 이론적 분석이 아닌 simulation 기반으로 적절한 compression rate를 구한 것이며, 더 다양한 simulation을 통해 보다 효율적인 compression rate를 얻을 수 있다.
%
%
%
%
%\subsubsection{Ciphertext size of NewHope}
%When $n=1024$, the ciphertext size of NewHope is  2176 bytes, when $n=512$, the ciphertext size of NewHope is  1088 bytes
%One of our main contribution(goal) is the reduction of ciphertext size.

\subsection{BCH Codes}
BCH codes are algebraic block codes widely used in digital communications and storage systems.
Unlike other advanced ECCs such as low-density parity-check (LDPC) codes and Turbo codes, BCH codes do not show an error floor problem and thus can achieve the required DFR ($\approx 2^{-140}$) \cite{Moon}, \cite{Wicker}.
Also, BCH codes can be theoretically analyzed whether the required DFR is achieved or not.
Therefore, BCH codes are ECCs suitable for PQC
In addition, various research activities \cite{LAC}, \cite{Fritzmann}, \cite{BCH constant}, \cite{ErrorDependency}, \cite{ErrorDependency2} are currently underway to apply BCH codes to the cryptosystem.

A BCH code is usually denoted by BCH($C_n$, $C_k$, $C_t$) where $C_n$ is the code length, $C_k$ is the dimension, and $C_t$ is the error-correction capability.
The code length $C_n$ of the primitive $C_q$-ary BCH code is $C_q ^ m-1, m = 3, 4, ... $, and the code length can be adjusted by applying many modification methods such as shortening and puncturing.
The dimension $C_k$ denotes the number of message symbols in a codeword and $C_t$ is the maximum number of errors that are always corrected \cite{Moon}, \cite{Wicker}.

%BCH code의 parameters로는 code length, dimension, correcting capability 로 구성된다. Primitive binary BCH code의 code length는 $C_n=2^m-1, m=3, 4, ...$이고, 때에 따라 shortening 기법을 적용하여 code length를 조정할 수  있다. The dimension $C_k$ of binary BCH code is related to the number of messages that can be transmitted. $C_t$ is the number of errors that can be corrected.

%BCH code는 대표적인 algbraic codes로써 다양한 장점들로 인해 digital communication과 storage system에서 널리 사용되고 있다. 대표적인 장점으로는 correcting capability를 설계할 수 있는 점이 있다. 이러한 장점으로 인해 다른 Error correcting codes such as LDPC code, Turbo code 와 달리 error floor 현상이 나타나지 않아 아주 낮은 DFR(BER)을 달성할 수 있다. 또는 채널에 대한 정확한 분석에 기반하면 설계된 BCH code의 DFR(BER) 성능에 대해 수학적으로 분석이 가능하다. 따라서 PQC에서 요구되는 DFR을 달성 여부를 시뮬레이션 기반이 아닌 수학적으로 판단할 수 있기 때문에 cryptosystem에 적용되기 적절한 error correcting code이다.

%BCH parameter 소개, shortening 소개

\section{Improving NewHope Based on Exact Noise Analysis and DFR Recalculation}

\subsection{An Information-theoretic View of NewHope}
\begin{figure}
\centering
\includegraphics[width=\textwidth]{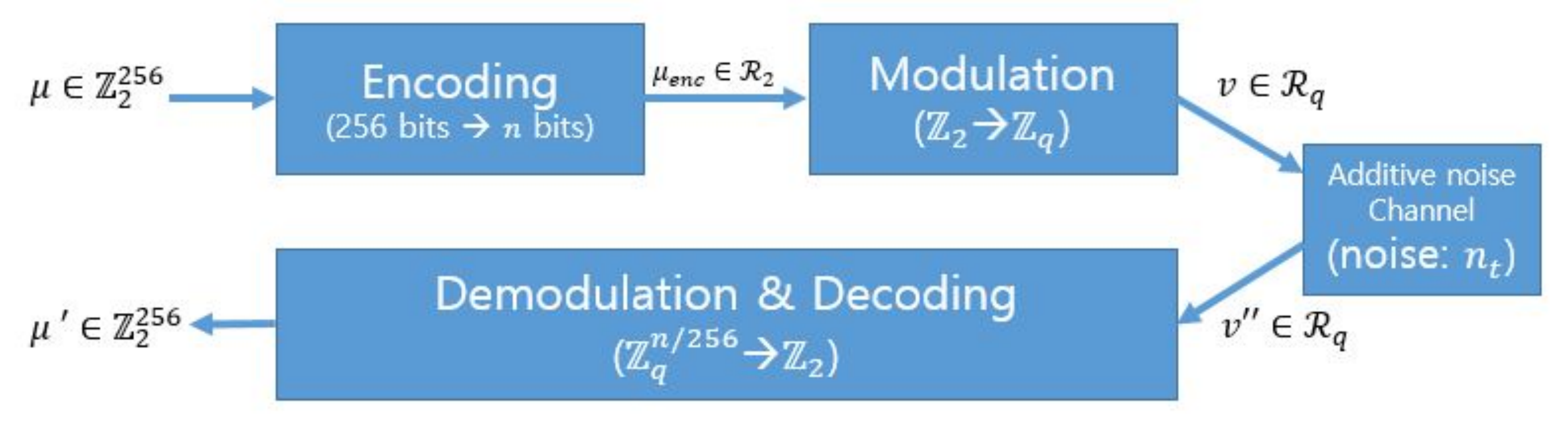}
\caption{An information-theoretic view of NewHope as a digital communication system $n=512$ or $n=1024$.} 
\label{Protocol0}
\end{figure}

In order to properly apply ECC to NewHope and facilitate analysis, it is necessary to understand the protocol of NewHope via an information-theoretic approach.
For NewHope, the mapping $\mathbb{Z}_2 \rightarrow \mathcal{R}_2$ and the mapping $\mathcal{R}_2 \rightarrow \mathbb{Z}_2 $ can be regarded as encoding and decoding of ECC, respectively. 
Also, the mapping $\mathcal{R}_2 \rightarrow \mathcal{R}_q$ and $\mathcal{R}_q \rightarrow \mathcal{R}_2$ can be regarded as modulation and demodulation, respectively.
Then NewHope can be understood as a digital communication system as follows.
Bob and Alice are transmitter and receiver, respectively, and the 256-bit shared secret key ($\mu$) is a message bit stream.
Also, the process of transmitting and receiving messages (Steps 4, 5, 6, and 7) can be viewed as a digital communication channel.
In more detail, the transmitter (Bob) generates a 256-bit message bit stream, encodes this massage, modulates the codeword bits to the symbols of $\mathbb{Z}_q$ and transmits the resulting signal (Step 4).
At the receiver (Alice), the received signal through the noisy channel is demodulated and decoded (Step 7).
For NewHope, a process of adding the compression noise and the difference noise generated in Steps 5 and 6 can be regarded as noisy communication channel.
This overall process in Steps 4-7 can be described as a digital communication system shown in Fig. \ref{Protocol0}.
% NewHope protocol을 digital communication system으로도 이해할 수 있다. Bob과 Alice를 각각 transmitter와 receiver로 볼 수 있고, 서로 공유하고자 하는 256-bit shared secret key를 message로 볼 수 있다. 또한 message를 주고 받는 과정들(step 5, 6, 7)을 channel로 볼 수 있다. 조금 더 자세히 설명하면, Transmitter에서는 message를 생성하고, ECC를 적용한 후 (z...) modulation하여 신호를 송신하는 과정으로 볼 수 있다 (Step 4). Reciever에서는 channel을 통해 수신된 신호를 demodulation한 후, 복호화하는 과정으로 볼 수 있다 (step 7). 그리고 channel은 step 5, 6을 겪으면서 발생되는 compression noise와 difference noise가 발생되어 더해지는 과정으로 볼 수 있다. 이러한 과정을 그림과 같이 표현할 수 있다.

In Fig. \ref{Protocol0}, $\mu_{enc}$ is the resulting signal after applying ATE to $\mu$, and $n_t$ represents the overall noise generated in Steps 5 and 6, which is called the total noise.
Since ATE simultaneously performs demodulation and decoding, demodulation and decoding are combined into one block.
After interpreting NewHope as a digital communication system, the DFR in NewHope is equivalent to the block error rate (BLER), i.e., $Pr(\mu \neq \mu')$, in a digital communication system.
Therefore, in order to accurately calculate the DFR of NewHope, exact analysis of the noisy channel is required.

%In Fig. \ref{Protocol0}, $C_n$ denotes the code length of ECC and $n_t$ represents noise that is generated in step 5 and 6. $\mu_{enc}$ is the result of ECC applied to $\mu$ and $\mu''_{enc}$ is result of $\mu_{enc}$ through the channel.

%In NewHope에서는 Encode에서는 4 repetition code가 적용되는 것이고, 따라서 C_n은 1024 and Modulation은 4 repetion이 적용된 $\mu_{enc}$의 element들에 0은 0으로 1은 \$lfloor \frac{q}{2} \rfloor$로 mapping하는 것이다.

% $C_n$ denotes 

%Exact noise analysis of NewHope is necessary to understanding NewHope in terms of ECC. Fig. \ref{NewHope in ECC} shows NewHope Protocol 

\subsection{Exact Noise Analysis of NewHope and DFR Recalculation}

%본 논문에서는 정확하게 difference noise와 compression noise의 distribution을 계산하여 최종적으로 Alice가 겪는 total noise의 distribution을 계산한 후, ATE가 적용된 DFR을 재계산한다.
%Comprssion noise는 
\subsubsection{Difference Noise, Compression Noise, and Total Noise Analysis}
Total noise $n_t$ is defined as the noise contained in the received 256-bit shared secret key before demodulation. 
Total noise of the $i$th coefficient $n_{t,i}$ of the polynomial $v''$ in Step 6 is represented as follows.

\begin{eqnarray}
n_{t,i} & = & (v''-v)_i \\
& =& (v'_{decomp}-us-v)_i \\
& =& (v'+n_c-us-v)_i \\
& =& (bs'+e''-ass'-e's)_i+n_{c,i} \\
& =& (es'-e's+e'')_i+n_{c,i} \\
& =& n_{d,i}+n_{c,i}
\label{total noise equation}
\end{eqnarray}
where $(\cdot)_i$ denotes the $i$th coefficient of polynomial, $n_c \in \mathcal{R}_q$ is the compression noise polynomial, $n_{c,i}$ is the $i$th coefficient of compression noise polynomial in $v''$, and $n_{d,i}$ is the $i$th coefficient of difference noise polynomial in $v''$.

\begin{figure}
\centering
\includegraphics[width=\textwidth]{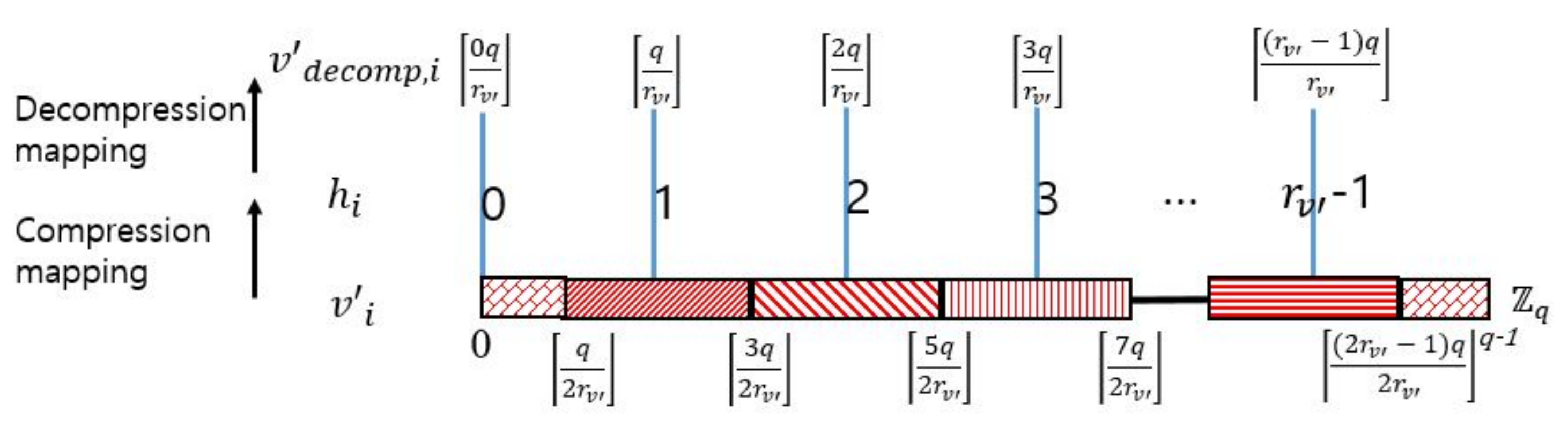}
\caption{Mapping corresponding to each of compression and decompression in NewHope.} 
\label{Compression}
\end{figure}

To analyze the compression noise $n_{c,i}$, we first need to investigate the coefficient of the polynomial $v'=ass '+ es' + e ''$ being compressed, where the coefficients of $s$, $s'$, $e$, and $e''$ follow the predetermined centered binomial distribution. 
However, since the coefficients of polynomial $a$ follow a uniform distribution, the coefficient of the compressed polynomial $h$ will eventually follow a uniform distribution.
A compression to $v'$ is performed by applying $\lfloor v'_i * r_{v'}/q \rceil$ to the coefficients $v'_i$ of $v'$ to obtain a polynomial $h$, where $\lfloor \cdot \rceil$ is rounding function that rounds to the closest integer, $r_{v'}$ denotes the compression rate on $v'$, and $r_{v'}=8$ for NewHope. 
Then the range of the compressed coefficients $h_i$ of $h$ is changed from $[0, q-1]$ to $[0, r_{v'} - 1]$ so that the number of bits required to store a coefficient is reduced from 14 bits ($=\lceil \log_2q \rceil$) for $v'$ to 3 bits ($=\lceil \log_2r_{v'} \rceil$) for $h$.
A decompression is performed by applying $\lfloor h_i * q/r_{v'} \rceil$ to each of the coefficients of $h$.
Then the coefficient takes the value from $ 0 $, $ \lfloor q / r_ {v'} \rceil $,$ \lfloor 2q / r_ {v'} \rceil $ \dots, $ \lfloor (r_{v'}-1)\cdot q / r_ {v '} \rceil $.
This compression and decompression are illustrated in Fig. \ref{Compression}, where the coefficients $v'_i$ of $ v '$ from different patterns (or ranges) are mapped to different $v_{decomp,i}$ values through compression and decompression.
In the end, compression and decompression can be seen as a rounding operation.
Therefore, the compression noise is inevitably generated with the maximum magnitude $\lfloor q/2r_{v'} \rfloor$ and the distribution of the compression noise is derived as follows:

\begin{equation}
n_c[x] = \left\{\begin{array}{ll}
q/r_{v'}, & 0 \le x \le \lceil \frac{q}{2r_{v'}} \rceil-1 \\
0, & \textrm{otherwise} \\
q/r_{v'} & q-2-\lceil \frac{q}{2r} \rceil \le x \le q-1 
\end{array} \right.
\end{equation}

%$r_{v'}$ denotes compression rate for the coefficients of $v'$, and $r_{v'}=8$ for NewHope. 이러면 $v'$의 coefficent들의 range가 [0,q-1]에서  [0,r_{v'}-1]로 변형되어, v'의 하나의 coefficient를 저장하기 위해 필요했던 14 bits가 3 bits로 감소한다. Decompression은 $h$의 coefficient들에 $\rceil h_i * q/r_{v'} \rfloor$를 적용하는 것으로 수행된다. 이 연산으로 인해 coefficient들은 [0,r_{v'}-1]에서 $0$, $\rceil q/r_{v'} \rfloor$, \dots, $\rceil  (r_{v'}-1)* q/r_{v'} \rfloor$로 변환된다. 따라서 compression/decompression으로 인해 noise가 많으면 $\lfoor q/2*r_{v'} \rfloor$만큼 필연적으로 발생하게 된다. 

In total noise, $n_{d,i}=(es'-e's + e'')_i$ is defined as the $i$th coefficient of the difference noise polynomial where the coefficients of $e$, $e'$, $e''$, $s$, and $s'$ follow the same centered binomial distribution.
In order to derive the distribution of difference noise, a number of convolution operations are required because the difference noise is a sum of many random variables, each of which is obtained by multiplying two random variables that follow the centered binomial distribution.
However, since it is difficult to calculate the multiple convolutions of the above distribution in closed form, the distribution of difference noise is numerically calculated \cite{Fritzmann}.

\begin{figure}
\centering
%\begin{subfigure}
\includegraphics[width=\textwidth]{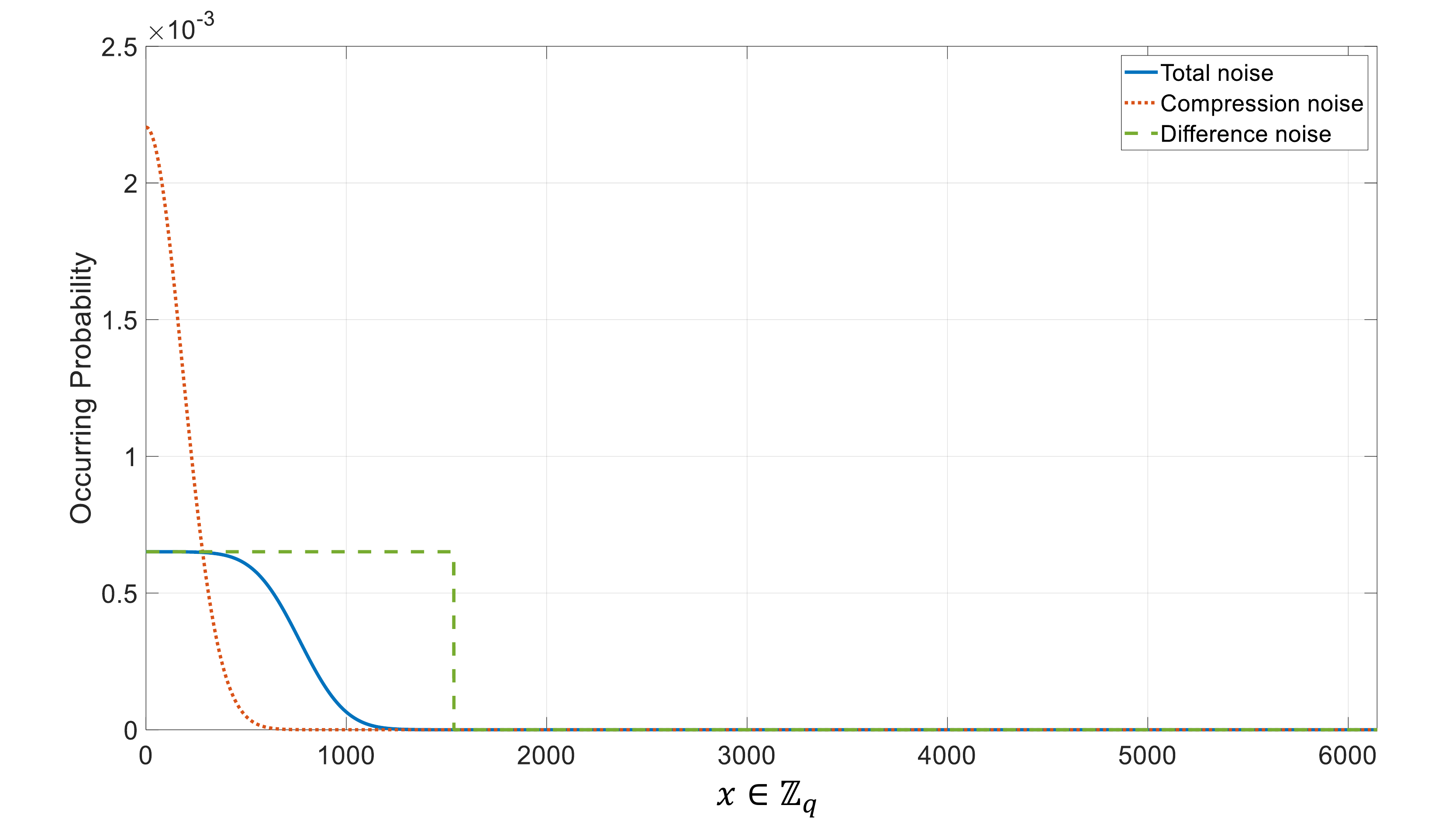}
\caption{Distributions of total noise, compression noise, and difference noise of NewHope (Distribution of total noise, compression noise, and difference noise is symmetric with respect to $\lfloor \frac{q}{2} \rfloor$ and $q=12289$). } 
\label{Total noise}
%\end{subfigure}

\end{figure}

Total noise is a sum of compression noise and difference noise which are independently generated.
Thus, the distribution of total noise is obtained by performing convolution of the distributions of compression noise and difference noise as shown in Fig. \ref{Total noise}.

%NewHope에는 포함되지 않았지만, \hat{u}에 compression을 수행한다면 total noise는 다음과 같이 다시 수정할 수 있다. \hat{u}의 $i$th 계수에 compression, decompression을 수행하면서 발생된 noise를 $n_{u,i}$라 하자.
% \hat{u}의 생성에 $a$가 포함되기 때문에, \hat{u}의 계수들은 uniform distribution을 따른다. 따라서 $v'$에 대한 compression noise와 마찬가지로 \hat{u}의 compression noise도 uniform distribution을 따른다.

If a process of compression on $\hat{u}$ is added to NewHope, total noise $n_{t,i}$ is modified as follows.

\begin{equation}
n_{t,i} =  n_{d,i}+n_{c,i}-(n_us)_i
\label{u compression}
\end{equation}
where $n_u\in \mathcal{R}_q$ is the compression noise polynomial generated while the compression and decompression to $\hat{u}$.

The coefficients of the polynomial $\hat{u}=\hat{a} \circ \hat{t} + NTT(e')$ follow a uniform distribution because the coefficients of $\hat{a}$ follow a uniform distribution.
A compression to $\hat{u}$ is performed by applying $\lfloor \hat{a} \cdot r_{\hat{u}}/q \rceil$ to the coefficients $\hat{u}_i$ of $\hat{u}$, where $r_{\hat{u}}$ denotes the compression rate on $\hat{u}$ and $\hat{u}=q$ for NewHope.

\begin{figure}
\centering
\includegraphics[width=\textwidth]{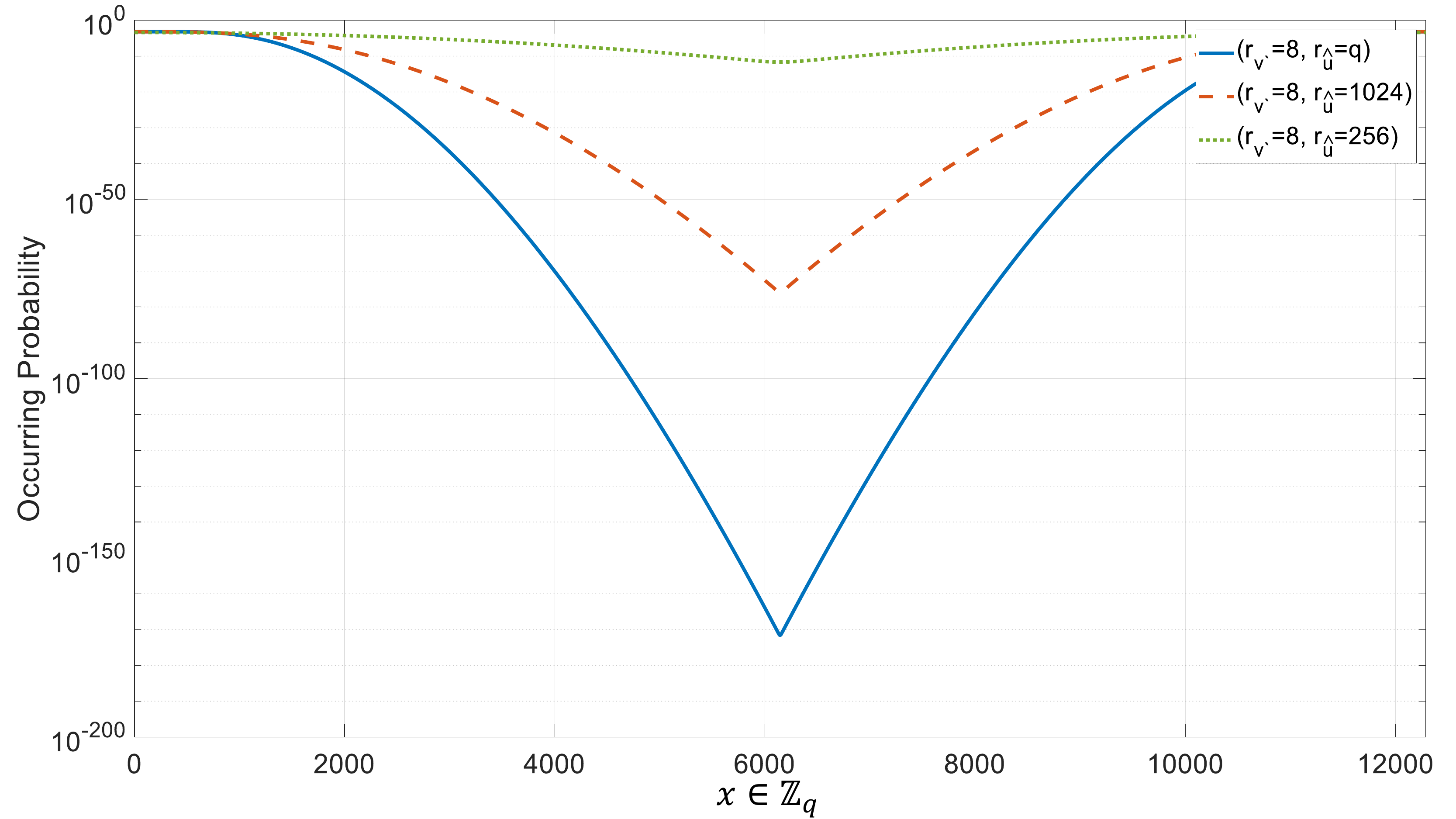}
\caption{Distributions of total noise for various compression rates on $v'$ and $\hat{u}$ ($n=1024$). } 
\label{U compression}
\end{figure}

Fig. \ref{U compression} shows the total noise distribution for the various compression rates of $r_{v'}$ and $r_{\hat{u}}$.
Note that the total compression noise of NewHope is obtained by ($r_{v'}=8$, $r_{\hat{u}}=q$).
$\hat{u}$ cannot be compressed as much as $v'$ because the compression noise of $\hat{u}$ is multiplied by $s$ to affect total noise as shown in (\ref{u compression}).

\subsubsection{DFR Recalculation}		
%It is difficult to calculate the exact DFR only with the total noise of one coefficient analyzed in the previous Section 3.2 because the coefficients of the product of the two polynomials are not independent of each other. 
%However, we assume that the coefficients are independent to compute the DFR.
%Because, as shown by the experimental results of [Fritzmann], there is almost no influence of DFR due to correlation of coefficients.
%Additionally, unlike the LAC, which is the target of [Error dependency], NewHope is expected to have little influence of DFR due to correlation of coefficients because $q$ and $k$ are larger.
%Nevertheless, since there is a little correlation between the coefficients, DFR criterion is set to $2^{-140}$ instead of $2^{-128}$ for margin.

The DFR of NewHope calculated in \cite{NewHope NIST} is defective because the compression noise $n_c$ is not considered and the centered binomial distribution is approximated by Gaussian distribution. 
In addition, ATE is not considered, which clearly affects the DFR of NewHope, and an upper bound of the DFR is derived using the Chernoff-Cramer bound instead of doing exact calculation \cite{NewHope NIST}, \cite{NewHope1}.
Also, instead of calculating $Pr(\mu' \neq \mu)$ of Fig. \ref{Protocol0} as DFR, $Pr(v_i \neq v''_i)$ of Fig. \ref{Protocol0} was calculated.
Therefore, current DFR values of NewHope, which are DFR $< 2 ^ {- 213}$ for $ n = 512 $ and DFR $< 2 ^ {- 216}$ for $ n = 1024$, are not correct and hence it is necessary to recalculate accurate DFR values of NewHope based on exact noise analysis. 

To recalculate the DFR of NewHope, it is assumed that the coefficients of polynomials are statistically independent to each other because it is shown by experiments in \cite{Fritzmann} that there is almost no influence on DFR by the correlation of coefficients.
Since it is typical to give a margin to the target DFR, the DFR requirement is usually set to $2^{-140}$ instead of $2^{-128}$.

\begin{figure}
\centering
\includegraphics[width=\textwidth]{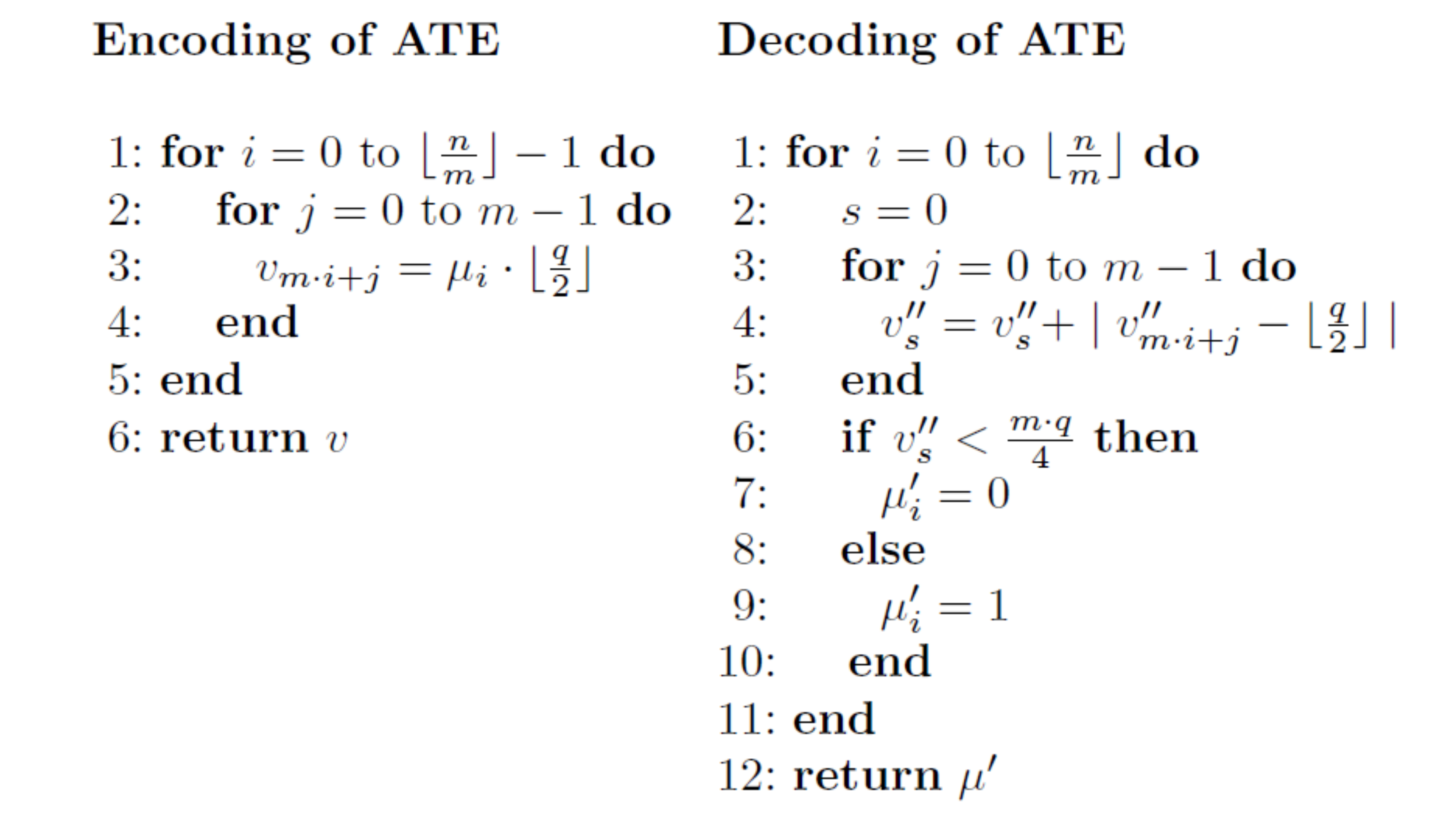}
\caption{Encoding and decoding of ATE for NewHope.} 
\label{ATE encoding decoding}
\end{figure}

In the previous section, the total noise $n_{t}$ is thoroughly analyzed.
In NewHope, ATE is used to encode and decode a message bit $ \mu_i $, and the encoding (including modulation) and decoding (including demodulation) procedures of ATE with $m$ repetitions are shown in Fig \ref{ATE encoding decoding} \cite{ATE}.
The encoding/modulation of ATE is performed such that one message bit $\mu_i$ is repeated $m$ times and each of them is mapped to the element of $\mathbb{Z}_q$ (usually either 0 or $\lfloor \frac{q}{2} \rfloor$) as the coefficients of $v$.
The demodulation/decoding of ATE sums up the $m$ absolute values of the differences of $\frac{q}{2}$ and the $m$ received coefficient values of $v''$ corresponding to the $m$ repeated coefficients.
Then, the sum $v''_s$ is compared with $m \cdot q / 4 $ to determine whether the message bit $\mu_i$ is 0 or 1 as follows.

\begin{equation}
v''_s \underset{0}{ \overset{1} \gtrless} \frac{m \cdot q}{4}
\end{equation}

Compared with the repetition code, ATE has the same encoding/modulation process, but uses different demodulation/decoding which is simple and effective because the square operation is not required for demodulation/decoding.
Therefore, although the error-correction capability of ATE is worse than those of other advanced ECCs, ATE still works well as an ECC for PQC such as NewHope.

%앞선 Section에서는 polynomial의 하나의 계수의 total noise를 분석하였다. NewHope에서는 하나의 message bit $\mu_i$가 demodulation/decoding 되기 위해서는 ATE를 통해 반복된 계수들이 필요하다. ATE의 encoding/moduation과 demodulation/decoding 그림X와 같다.
%ATE의 encoding/moduation은 하나의 message bit가 m번 반복되어 v의 coefficient들로 mapping 된다. 그리고 ATE의 demodulation/decoding은 m번 반복된 v의 계수들의 절댓값들을 합한 후 $\frac{m \cdot q}{4}$와 비교하여 message bit를 0 또는 1로 판정한다.
%ATE은 repetition code와 비교화여 encoding/modulation는 동일하지만, demodulation/decoding은 간단하지만 효과적이다. 따라서, 비록 다른 진보된 ECCs에 비해 오류 정정 능력은 낮지만, PQC에 적합한 ECC이다.

$Pr(\mu_i' \neq \mu_i)$ is called bit error rate (BER) of $\mu_i$ for $i=0, 1, \dots, 255$ as follow.
First, the error distribution $P_{v''}$ of $v''_i$ is calculate as follow.  

\begin{eqnarray}
P_{v''} &=& Pr \Big( \left| v''_i-\lfloor \frac{q}{2} \rfloor \right| < \frac{m \cdot q}{4} \mid v_i=\lfloor \frac{q}{2} \rfloor \Big)Pr\Big(v_i=\lfloor \frac{q}{2} \rfloor \Big)\\
& &+ Pr \Big( \left| v''_i-\lfloor \frac{q}{2} \rfloor \right| \geq \frac{m \cdot q}{4} \mid v_i=0 \Big)Pr\Big(v_i=0\Big)
\end{eqnarray}
Second, the error distribution $P_{\mu'}$ of $\mu_i$ is calculated as follow.

\begin{equation}
P_{\mu'}=\underbrace{P_{v''} \otimes P_{v''} \otimes \cdots \otimes P_{v''}}_{m}
\end{equation}
where $\otimes$ is the convolution operation and the range of $P_{\mu'}$ is $[0,m \cdot q -1]$.
Let $BER_m$ be BER when the ATE with $m$ repetition is used.
$BER_m$ is calculated as follow.

\begin{equation}
BER_m = P_{\mu'}\Big(x < \frac{m \cdot q}{4}\Big)Pr(\mu_i=1) + P_{\mu'}\Big(x \geq \frac{m \cdot q}{4}\Big)Pr(\mu_i=0)
\label{BERm}
\end{equation}
where $Pr(\mu_i=0)=Pr(\mu_i=1)=\frac{1}{2}$.

Since we assume that the coefficients of polynomial $v''$ are independent, $Pr(\mu_i' \neq \mu_i)=Pr(\mu_j' \neq \mu_j)$ for $i, j=0,1,\dots ,n-1$.
Therefore, BLER, which is in fact the DFR, of ATE with $m$ repetitions in NewHope is calculated as $1-(1-BER_m)^{256}$.

Based on the above analysis, the DFR of NewHope is recalculated as $2^{-474}$ and $2^{-431}$ for $n=1024$ and $n=512$, respectively.
These DFR values are much smaller than the corresponding DFR values $2^{-216}$  and $2^{-213}$ from the NewHope specification \cite{NewHope NIST}.
Therefore, it is possible to improve the security and bandwidth efficiency of NewHope only by utilizing this DFR margin.

\subsection{Improved Security and Bandwidth Efficiency of NewHope Based on Recalculated DFR}

\subsubsection{Improved Security}
Since there exists a trade-off relation between security level and DFR, it is necessary to properly select the noise parameter $k$ of centered binomial distribution such that the security level and the DFR are appropriately determined.
Even if NewHope is designed to have an extremely low DFR, the security level can be more improved by using new  DFR margin obtained from recalculated DFR.

% exact noise analysis를 통해 예상되었던 NewHope의 DFR이 훨씬 좋은 성능을 나타낸 것을 확인하였다. DFR은 NewHope parameters 중에서 noise variance와 가장 밀접한 관계가 있다. 
% DFR과 security는 서로 trade-off 관계에 있다. Security level을 증가시키면 암호 안정성이 증가하지만 이에 따라 DFR이 증가하여 Alice와 Bob이 원활하게 secret key를 주고 받을 수 없다. 반대로 지나치게 원할하게 Alice와 Bob이 secret key를 주고 받는다면 security level이 감소하여 암호 안정성이 낮아지게 된다. 따라서 parameter를 잘 정하여 security level과 DFR을 적절히 정해지도록 설정해야한다. 그러나 NewHope는 지나치게 낮은 DFR 성능을 갖도록 설계되어 있기 때문에 DFR 성능을 조금 포기하더라도 security level을 증가시킬 수 있다. 이렇게 하는 이유는 암호 안정성이 DFR 성능보다 우선되는 성능이기 때문이다. 
% 그러나 지나치게 우수한 DFR 성능을 나타내기 때문에 이 점을 활용하여 security level을 높힐 수 있다. 

\begin{table}[]
\centering
\caption{Improved security level of NewHope based on new DFR margin (noise parameter $k=8$ for current NewHope).}
\begin{tabular}{|c|c|c|c|c|}
\hline
\multicolumn{1}{|c|}{$n$}  & \multicolumn{1}{|c|}{$k$} & \multicolumn{1}{|c|}{DFR}         & \begin{tabular}[c]{@{}c@{}}Cost of primal attack\\ Classical/Quantum [bits]\end{tabular} & \begin{tabular}[c]{@{}c@{}}Cost of dual attack\\ Classical/Quantum [bits]\end{tabular} \\ \hline
\multirow{2}{*}{1024} & 8   & $ \approx 2^{-474}$ & \multicolumn{1}{|c|}{259/235}                                                            & \multicolumn{1}{|c|}{257/233}                                                         
\\ \cline{2-5}
% & 14  & $\approx2^{-165}$ & \multicolumn{1}{|c|}{278/252}                                                            & \multicolumn{1}{|c|}{276/250}                                                          \\ \cline{2-5}
 & 15  & $\approx2^{-145}$ & \multicolumn{1}{|c|}{280/254}                                                            & \multicolumn{1}{|c|}{279/253}                                                          \\ \hline
% & 16  & $\approx2^{-128}$ & \multicolumn{1}{|c|}{282/256}                                                            & \multicolumn{1}{|c|}{281/255}                                                          \\ \hline
\multirow{2}{*}{512}  & 8  & $\approx2^{-431}$ & \multicolumn{1}{|c|}{112/101}                                                            & \multicolumn{1}{|c|}{112/101}                                                          \\ \cline{2-5}
%  & 13  & $\approx2^{-176}$ & \multicolumn{1}{|c|}{120/109}                                                            & \multicolumn{1}{|c|}{119/108}                                                          \\ \cline{2-5}
  & 14  & $\approx2^{-154}$ & \multicolumn{1}{|c|}{121/110}                                                            & \multicolumn{1}{|c|}{121/110}                                                          \\ \hline
%  & 15  & $\approx2^{-135}$ & \multicolumn{1}{|c|}{122/111}                                                            & \multicolumn{1}{|c|}{122/111}                                                          \\ \hline
\end{tabular}
\label{NewHope Security level}
\end{table}

Table \ref{NewHope Security level} shows the improved security levels which are calculated by assuming cost of the primal attack and dual attack \cite{Attack} to NewHope. 
It is possible to improve the security level by 8.0 \% ($n=1024$, $k=15$) and 8.9 \% ($n=512$, $k=14$) while keeping the required DFR of $2^{-140}$ compared with the current NewHope.
Note that such security level improvement does not require too much increase of time/space complexity in NewHope because it only changes the noise parameter $k$ without any additional process.

% A의 DFR을 충분히 달성하면서도 security level을 8.0 \%와 8.9 \%를 개선시킬 수 있다. 그리고 noise variance parameter만 8에서 15, 8에서 14로 변경하면 되기 때문에 protocol의 복잡도 증가없이 security level을 개선시킬 수 있다.

%Primal attack과 Dual attack를 고려하였고, 둘중에 낮은 security level을 기준으로 삼았다.

\subsubsection{Improved Bandwidth Efficiency}
The bandwidth efficiency of NewHope can also be improved by utilizing new DFR margin. 
% Ciphertext size를 감소시킬수록 compression noise 증가로 인해 DFR 성능이 열화된다. 
An improvement of bandwidth efficiency is achieved by reducing the ciphertext size which, however, increases the compression noise resulting in the DFR degradation.
However, even with such increased compression noise, both the improvement of bandwidth efficiency and the required DFR of $2^{-140}$ can be achieved by utilizing DFR margin.

%그렇지만 compression noise가 증가되어도 DFR margin을 통해 DFR을 달성할 수 있다.

\begin{table}[]
\caption{Improved bandwidth efficiency of NewHope based on new DFR margin.}
\centering
\begin{tabular}{|c|l|l|l|}
\hline
\multicolumn{1}{|c|}{$n$}  & \multicolumn{1}{c|}{($r_{v'}$, $r_{\hat{u}}$)}        & \multicolumn{1}{|c|}{Ciphertext reduction (\%)} & \multicolumn{1}{|c|}{DFR}         \\ \hline
\multirow{3}{*}{1024} & \multicolumn{1}{c|}{(8,$q$)} & \multicolumn{1}{c|}{0 (Current NewHope)} & $\approx 2^{-474}$ \\ \cline{2-4}
 & \multicolumn{1}{c|}{(4,$q$)}    & \multicolumn{1}{c|}{5.9}               & $\approx 2^{-227}$ \\ \cline{2-4}
 & \multicolumn{1}{c|}{(8, 1024)} & \multicolumn{1}{c|}{23.5}              & $\approx2^{-199}$ \\ \hline
\multirow{4}{*}{512}    & \multicolumn{1}{c|}{(8,$q$)}    & \multicolumn{1}{c|}{0 (Current NewHope)}               & $\approx2^{-431}$ \\ \cline{2-4}

  & \multicolumn{1}{c|}{(4,$q$)}    & \multicolumn{1}{c|}{5.9}               & $\approx2^{-420}$ \\ \cline{2-4}
  & \multicolumn{1}{c|}{(4, 2048)}  & \multicolumn{1}{c|}{23.5}              & $\approx2^{-155}$  \\ \cline{2-4}
  & \multicolumn{1}{c|}{(8, 1024)}  & \multicolumn{1}{c|}{23.5}              & $\approx2^{-185}$  \\ \hline
\end{tabular}
\label{NewHope Bandwidth efficiency}
\end{table}

Table \ref{NewHope Bandwidth efficiency} shows the improved bandwidth efficiency of NewHope achieved by  additional ciphertext compression. 
The results in Table \ref{NewHope Bandwidth efficiency} are obtained in two ways.
The first way ($r_{v'}=4$, $r_{\hat{u}}=q$) changes only the compression rate on $ v '$ from 8 (3 bits per coefficient) to 4 (2 bits per coefficient) to improve the bandwidth efficiency, which does not change the protocol of NewHope and hence does not increase the time/space complexity.
The second way (the remaining results) is to improve the bandwidth efficiency by doing additional compression on both $ \hat{u} $ and $ v '$, which results in a slight increase of time/space complexity  because compression of $ \hat {u} $ is added to the NewHope protocol.
However, the second way shows about four times improvement in bandwidth efficiency over the first way while keeping the target DFR.

%n=512와 n=1024일때 $v'$에 대해서만 추가적인 compression을 수행하여 bandwidth efficiency를 개선시킬 수 있다. while achieving the DFR of  $2^{-140}$ sufficiently. 이 경우는 
%Table X의 결과를 크게 두가지로 나눌 수 있다. 첫번째로는 $v'$에 대해서만 compression rate를 3에서 2로 감소시켜 bandwidth efficiency 를 개선시킨다. 이러한 방법은 NewHope의 protocol에 곧바로 적용이 가능하면 어떠한 복잡도 증가가 없다. 두번째로는 $v'$ 뿐만 아니라 $\hat{u}$에 대해서도 추가적인 compression을 수행하여 bandwidth efficiency를 개선시키는 것이다. 이 방법은 NewHope의 protocol에 없는 $\hat{u}$의 compression 과정이 추가되어야 하기 때문에 복잡도 증가가 동반된다. 그러나 첫번째 방법에 비해 4배의 bandwidth efficiency 개선이 있기 때문에 추후 NewHope에 고려할만한 가치가 있다.

\section{Improving Security and Bandwidth Efficiency of NewHope Using Error-correction Schemes}
% ECC 의 기본 개념
ECCs are used to format the transmitted information so as to protect the information from the noisy channel. 
Such protection is obtained by adding systematic redundancy to the information, which enables the receiver to detect and possibly correct errors. 
Therefore, ECCs are an essential part of digital communication/storage systems.

ATE is used in NewHope as an ECC, which was proposed by P{\"o}ppelmann and G{\"u}neysu \cite{ATE}
However, since ATE is based on the repetition code, it shows low error-correction capability.
Therefore, by applying advanced ECC to NewHope, the security and bandwidth efficiency can be significantly improved.

%ATE는 repetition code와 유사하며, encoding/modulation이 동시에 가능하고, decoding/demodulation도 역시 동시에 가능하여 구현에 아주 용이하다. 그러나 repetition code를 기반이기 때문에 redundancy를 효율적으로 활용하지 못하여 낮은 오류 정정 능력을 갖는 단점이 있다. 따라서 NewHope에 진보된 ECC를 적용한다면 security와 bandwidth efficiency를 상당히 개선시킬 수 있다.
There are various good-performing ECCs such as turbo codes, LDPC codes, and BCH codes.
Turbo codes provide good error-correction performance close to the Shannon limit and the implementation of encoder is rather simple.
LDPC codes also show good error-correction performance and enable high-speed processing.
However, it is difficult to estimate low BLER of turbo codes and LDPC codes because they commonly have an error floor problem and no analytic way to accurately estimate very low BLER.
In contrast, BCH codes do not show error-correction performance close to the Shannon limit but low BLER can be analyzed by calculating an upper bound of BLER.
Also, BCH codes do not show an error floor problem.
Therefore, BCH codes can be a better choice for PQC because a very low DFR is required for PQC.
In this section, we investigate concatenated coding schemes of ATE and BCH code to combine the advantages of ATE and BCH code. 
However, the use of BCH codes is accompanied by an increase of implementation complexity and a possibility of side channel attacks.
Nevertheless we focus on theoretically investigating how much the security and bandwidth efficiency of NewHope can be improved by using ECCs rather than dealing with the implementation issues.
Note that implementation issues such as constant-time ECC \cite{BCH constant} and robustness to the side channel attacks are actively studied research topics.

\begin{figure} [h]
\centering
\includegraphics[width=\textwidth]{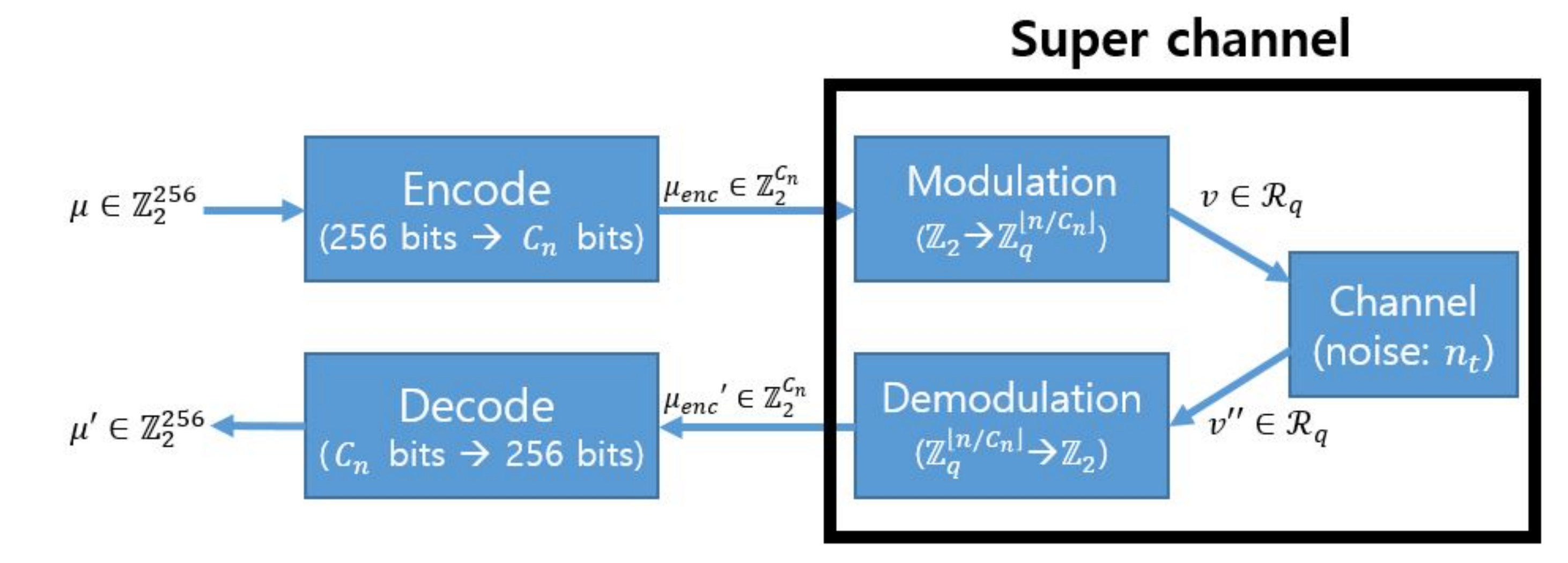}
\caption{An information-theoretic view of NewHope in terms of ECC and super channel.} 
\label{SuperChannel}
\end{figure}

From the viewpoint of ECC, the output of encoder and the input into decoder should be well defined, and hence the modulation, channel, and demodulation can be regarded as a super channel between them as shown in Fig. \ref{SuperChannel}.
Note that this super channel is a binary symmetric channel (BSC).
Then we can easily calculate the BLER (DFR) of NewHope with ECC after we obtain the cross-over probability of super channel by applying the total noise analysis.

\subsection{Candidates of Error-correction Schemes}
After analyzing various concatenated coding schemes for NewHope, we select the following four options (or ECCs) to validate the role of ECCs in NewHope.
Note that these four options do not mean the best ECCs but they can show that NewHope or other lattice-based cryptosystems can be dramatically improved in terms of security, bandwidth, and DFR by using ECCs even if there are still some implementation problems to be solved.
%Transmitter는 256-bits message를 1024 symbols로 만들어주면 된다. 따라서 ATE
%Foure options에 적용되는 BCH code로는 dimension이 256과 같거나 유사하고, code length가 $\lfloor \frac{m}{n} \rfloor$ for $m=$1, 2, and 3의 각각에 같거나 유사한 것들 중에서 narrow-sense primitive BCH code를 첫번째로 고려하였고, 두번째는 shortened BCH code를 고려하였다.
As the BCH codes used in four options, first, narrow-sense primitive BCH codes having the dimension equal to or similar to 256 and the code length equal to or similar to $ \lfloor \frac {n} {m} \rfloor $ for $ m = $ 1, 2, and 3 are considered.
Second, shortened BCH codes satisfying the above conditions are considered.

For all four options, BCH code is used as an outer code outside the super channel and ATE is used as an inner code for encoding/modulation and decoding/demodulation inside the super channel as given in Fig. \ref{SuperChannel}.

% BCH code는 super channel 밖에서 encoding과 decoding을 하는 outer code로써 활용되고, ATE는 super channel 안에서 encoding/modulation 과 decoding/demodulation을 하는 inner code로써 활용된다. 

- Option 1\\
Inner code: ATE(3-Repetition) + Outer code: BCH(341,260,9)\\
Parameters: $C_n=341$, $m=\lfloor \frac{n}{C_n} \rfloor = 3$ \\
A 256-bit message $\mu$ is encoded by BCH (341, 260, 9) which is the shortened code of BCH(511,430,9).
This encoder requires 260 bits as its input, while the size of message in NewHope is 256 bits.
Therefore, the input into the encoder is generated by padding four zeros to the 256-bit message $\mu$.
Then, the 341-bit BCH codeword becomes the input into the ATE.
The 1023-bit output of ATE ($m=3$) and the additional zero are used as 1024 coefficients to generate $v$.
With Option 1, up to 9 bit errors are corrected.

- Option 2\\
Inner code: ATE(2-Repetition) + Outer code: BCH(511,259,30)\\
Parameters: $C_n=511$, $m=\lfloor \frac{n}{C_n} \rfloor = 2$ \\
BCH(511,259,30) is a narrow-sense primitive BCH code and by padding three zeros to the 256-bit message $\mu$, the input into the BCH encoder is generated.
Also, the input of ATE encoder is generated by padding one zero to the 511-bit BCH codeword. 
With Option 2, up to 30 bit errors are corrected.

- Option 3\\
Inner code: ATE(1-Repetition) + Outer code: BCH(1023,258,106)\\
Parameters: $C_n=1023$, $m=\lfloor \frac{n}{C_n} \rfloor = 1$ \\
BCH(1023,258,106) is a narrow-sense primitive BCH code and by padding two zeros to the 256-bit message $\mu$, the input into the BCH encoder is generated.
Also, the input of ATE encoder is generated by padding one zero to the 1023-bit BCH codeword. 
With Option 3, up to 106 bit errors are corrected.

- Option 4\\
Inner code: ATE(1-Repetition) + Outer code: BCH(511,259,30)\\
Parameters: $C_n=511$, $m=\lfloor \frac{n}{C_n} \rfloor = 1$ \\
Option 4 is used for NewHope with $n=512$.
By adding three zeros to the 256-bit message $\mu$, the input into the BCH encoder is generated.
Also, the input of ATE encoder is generated by padding one zero to the 511-bit BCH codeword.
With Option 4, up to 30 bit errors are corrected.

\begin{table}[h]
\centering
\caption{Cross-over probability of super channel for Options 1, 2, 3, and 4.}
\begin{tabular}{|l|l|l|}
\hline
\multicolumn{1}{|c|}{Option} & \multicolumn{1}{c|}{ATE} & \multicolumn{1}{c|}{Cross-over probability of super channel} \\ \hline
NewHope                     & \multicolumn{1}{c|}{$m=4$}            & \multicolumn{1}{c|}{$1.3277 \times 10^{-145}$}                                      \\ \hline
Option 1                     & \multicolumn{1}{c|}{$m=3$}            & \multicolumn{1}{c|}{$8.3884 \times 10^{-110}$}                                      \\ \hline
Option 2                     & \multicolumn{1}{c|}{$m=2$}            & \multicolumn{1}{c|}{$6.0045 \times 10^{-74}$}    \\ \hline
Option 3                     & \multicolumn{1}{c|}{$m=1$}             & \multicolumn{1}{c|}{$5.1119 \times 10^{-38}$}                                       \\ \hline
Option 4                     & \multicolumn{1}{c|}{$m=1$}             & \multicolumn{1}{c|}{$1.7993 \times 10^{-67}$} \\ \hline
\end{tabular}
\label{cross-over prob. of super channel}
\end{table}

The repetition number $m$ of ATE affects the cross-over probability of super channel in Fig. \ref{SuperChannel}.
Obviously, the larger the repetition number $m$ of ATE, the lower the cross-over probability of super channel. 
Table \ref{cross-over prob. of super channel} shows the cross-over probability of super channel according to the repetition number of ATE, which is calculated by using (\ref{BERm}) with $m=$1, 2, 3, and 4.
However, as the repetition number $m$ of ATE increases, the code rate of BCH code also increases and hence the error-correction capability of BCH code degrades.
Note that for NewHope with $m=4$, the cross-over probability is the same as the DFR because there is no outer code.
In summary, there exists a trade-off between the repetition number $m$ of ATE (i.e., the quality of super channel) and the error-correction capability of BCH code (outer code).
Therefore, it is very important to determine the optimum repetition number $m$ of ATE and the error-correction capability of BCH code by investigating this trade-off.

In general, decoding of BCH codes is performed by the Berlekamp-Massey (BM) algorithm and the decoding complexity increases linearly with the error-correction capability $C_t$.
Therefore, in terms of implementation complexity Option 1 is the best choice because it has the lowest decoding complexity among four options.

% ATE의 repetition 수가 달라질 수

\subsection{Bandwidth Efficiency and DFR for Various Compression Rates on $v'$ and $\hat{u}$}

We consider various compression rates ($r_{v'}$,$r_{\hat{u}}$) on $v'$ and $\hat{u}$ as given in Table \ref{Compression rates table}.
The reason for choosing these compression rates is that the target DFR $2^{-140}$ for NewHope cannot be achieved only by using ATE due to the increased compression noise as can be seen from Table \ref{Compression rates table}.
Table \ref{Compression rates table} also shows the ciphertext size and DFR when each compression rate is applied to NewHope where the ciphertext size is calculated as $\frac{n}{8}(\lceil \log_2r_{v'} \rceil + \lceil \log_2r_{\hat{u}} \rceil)$.
Therefore, we can check if the target DFR can be achieved by applying Options 1, 2, 3, and 4 to NewHope with these compression rates.
Note that the ciphertext size reduction is related to compression noise, so it affects DFR and does not affect security.
%Table \ref{Compression rates table}에는 다양한 compression rate들이 NewHope에 적용되었을때의 ciphertext의 size와 DFR의 결과들이 나타나있다.
%이러한 compression rate들을 고려한 이유는 Table \ref{Compression rates table}에서 알 수 있듯이 NewHope에 적용된 ATE로는 NIST의 DFR 기준을 만족시키지 못한다. 
%그러나 이러한 compression rate들에 대해서도 ATE보다 강력한 ECC를 적용한다면 충분히 NIST의 DFR 기준을 만족시킬 수 있다. 

\begin{table}[h]
\centering
\caption{Various compression rates on $v'$ and $\hat{u}$ and the corresponding size of ciphertext and DFR ((8,$q$) is the compression rate of current NewHope).}
\begin{tabular}{|c|l|l|l|}
\hline
\multicolumn{1}{|c|}{$n$}   & \begin{tabular}[c]{@{}l@{}}($r_{v'}$, $r_{\hat{u}}$)\end{tabular} & Ciphertext size (bytes) & \multicolumn{1}{c|}{DFR}       \\ \hline
\multirow{7}{*}{1024} & \multicolumn{1}{c|}{(8,$q$)} & \multicolumn{1}{c|}{2176}   & $\approx 2^{-474}$ \\ \cline{2-4}
& \multicolumn{1}{c|}{(8,512)} & \multicolumn{1}{c|}{1536}   & $\approx 2^{-75}$ \\ \cline{2-4}
                       & \multicolumn{1}{c|}{(8,256)} & \multicolumn{1}{c|}{1408}   & $\approx 2^{-20}$ \\ \cline{2-4}
& \multicolumn{1}{c|}{(8,128)}                                                       & \multicolumn{1}{c|}{1280}          & $\approx 2^{-1}$  \\ \cline{2-4}
& \multicolumn{1}{c|}{(4,512)}                                                       & \multicolumn{1}{c|}{1408}          & $\approx 2^{-99}$ \\ \cline{2-4}
& \multicolumn{1}{c|}{(4,256)}                                                       & \multicolumn{1}{c|}{1280}          & $\approx 2^{-11}$ \\ \cline{2-4}
& \multicolumn{1}{c|}{(4,128)}                                                       & \multicolumn{1}{c|}{1152}          & $\approx 2^{-1}$  \\ \cline{2-4} \hline
\multirow{4}{*}{512}  & \multicolumn{1}{c|}{(8,$q$)}                                                       & \multicolumn{1}{c|}{1088}           &   $\approx 2^{-431}$        \\ \cline{2-4}
& \multicolumn{1}{c|}{(8,512)}                                                       & \multicolumn{1}{c|}{768}           &   $\approx 2^{-33}$        \\ \cline{2-4}
  & \multicolumn{1}{c|}{(8,256)}                                                       & \multicolumn{1}{c|}{704}  &    $\approx 2^{-7}$       \\ \cline{2-4}
  & \multicolumn{1}{c|}{(4,1024)}                                                      & \multicolumn{1}{c|}{768}  &    $\approx 2^{-43}$       \\ \cline{2-4}
  & \multicolumn{1}{c|}{(4,512)}                                                       & \multicolumn{1}{c|}{704}  &    $\approx 2^{-15}$       \\ \cline{2-4} \hline
\end{tabular}
\label{Compression rates table}
\end{table}

If a BCH($C_n$, $C_k$, $C_t$) is used, the DFR (or BLER) of NewHope over the super channel (or BSC) in Fig. \ref{SuperChannel} with the cross-over probability $p$ is calculated as 

\begin{equation}
DFR = 1 - \sum_{i=0}^{C_t} {C_n \choose i}p^{i}(1-p)^{C_n-i}.
\label{BCH DFR}
\end{equation}

\begin{table}[h]
\centering
\caption{DFR and ciphertext size of NewHope when Options 1, 2, and 3 are used with various compression rate when $n=1024$.}
\begin{tabular}{|c|l|l|c|}
\hline
($r_{v'}$, $r_{\hat{u}}$)& \multicolumn{1}{c|}{ECC Option}        & \multicolumn{1}{c|}{DFR}   & \begin{tabular}[c]{@{}c@{}}Ciphertext size\\ (reduction rate)\end{tabular}      \\ \hline

\multirow{4}{*}{\begin{tabular}[c]{@{}l@{}}(8,512)\end{tabular}} & Only ATE (NewHope) & $2^{-75}$ &\multirow{4}{*}{\begin{tabular}[c]{@{}l@{}}\multicolumn{1}{c}{1536 bytes} \\ (29.4\%)\end{tabular}}  \\ \cline{2-3} 
                                                                                                      & Option 1        & $2^{-569}$ & \\ \cline{2-3} 
                                                                                                      & Option 2        & $2^{-1177}$ & \\ \cline{2-3} 
                                                                                                      & Option 3        & $2^{-2000}$ & \\ \hline \hline

\multirow{4}{*}{\begin{tabular}[c]{@{}l@{}}(8,256)\end{tabular}} & Only ATE (NewHope) & $2^{-20}$ &\multirow{4}{*}{\begin{tabular}[c]{@{}l@{}}\multicolumn{1}{c}{1408 bytes} \\ (35.3\%)\end{tabular}}   \\ \cline{2-3} 
                                                                                                      & Option 1        & $2^{-151}$ &  \\ \cline{2-3} 
                                                                                                      & Option 2        & $2^{-317}$ & \\ \cline{2-3} 
                                                                                                      & Option 3        & $2^{-467}$ & \\ \hline \hline

\multirow{4}{*}{\begin{tabular}[c]{@{}l@{}}(8,128)\end{tabular}} & Only ATE (NewHope) & $2^{-1}$ &\multirow{4}{*}{\begin{tabular}[c]{@{}l@{}}\multicolumn{1}{c}{1280 bytes} \\ (41.2\%)\end{tabular}}   \\ \cline{2-3} 
                                                                                                      & Option 1        & $2^{-5}$ & \\ \cline{2-3} 
                                                                                                      & Option 2        & $2^{-8}$ & \\ \cline{2-3} 
                                                                                                      & Option 3        & $2^{-2}$ & \\ \hline \hline

\multirow{4}{*}{\begin{tabular}[c]{@{}l@{}}(4,512)\end{tabular}} & Only ATE (NewHope) & $2^{-40}$ &\multirow{4}{*}{\begin{tabular}[c]{@{}l@{}}\multicolumn{1}{c}{1408 bytes} \\ (35.3\%)\end{tabular}}   \\ \cline{2-3} 
                                                                                                      & Option 1        & $2^{-302}$ & \\ \cline{2-3} 
                                                                                                      & Option 2        & $2^{-620}$ & \\ \cline{2-3} 
                                                                                                      & Option 3        & $2^{-1016}$ & \\ \hline \hline

\multirow{4}{*}{\begin{tabular}[c]{@{}l@{}}(4,256)\end{tabular}} & Only ATE (NewHope) & $2^{-11}$ &\multirow{4}{*}{\begin{tabular}[c]{@{}l@{}}\multicolumn{1}{c}{1280 bytes} \\ (41.2\%)\end{tabular}}   \\ \cline{2-3} 
                                                                                                      & Option 1        & $2^{-85}$ & \\ \cline{2-3} 
                                                                                                      & Option 2        & $2^{-168}$ & \\ \cline{2-3} 
                                                                                                      & Option 3        & $2^{-222}$ & \\ \hline \hline

\multirow{4}{*}{\begin{tabular}[c]{@{}l@{}}(4,128)\end{tabular}} & Only ATE (NewHope) & $2^{-1}$ &\multirow{4}{*}{\begin{tabular}[c]{@{}l@{}}\multicolumn{1}{c}{1152 bytes} \\ (47.1\%)\end{tabular}}   \\ \cline{2-3} 
                                                                                                      & Option 1        & $2^{-1}$ & \\ \cline{2-3} 
                                                                                                      & Option 2        & $2^{-1}$ & \\ \cline{2-3} 
                                                                                                      & Option 3        & $2^{-0}$ & \\ \hline 

%\multirow{4}{*}{\begin{tabular}[c]{@{}l@{}}(2,$q$)\end{tabular}} & Only ATE (NewHope) & $2^{-9}$ &\multirow{4}{*}{\begin{tabular}[c]{@{}l@{}}Ciphertext \\ \\ 11.8 \% reduction\end{tabular}}   \\ \cline{2-3} 
%                                                                                                      & Option 1        & $2^{-65}$ & \\ \cline{2-3} 
%                                                                                                      & Option 2        & $2^{-120}$ & \\ \cline{2-3} 
%                                                                                                      & Option 3        & $2^{-120}$ & \\ \hline 
\end{tabular}
\label{Bandwidth efficiency improvement using ECC}
\end{table}

Table \ref{Bandwidth efficiency improvement using ECC} shows the DFR and ciphertext size (or improvement of bandwidth efficiency) reduction of NewHope with $n=1024$ when Options 1, 2, and 3 are used with various compression rates given in Table \ref{Compression rates table}.
The DFR is calculated by using (\ref{BCH DFR}) and the cross-over probability of super channel in Table \ref{cross-over prob. of super channel}.
The target DFR can be achieved with all the Options 1, 2, and 3 if the coefficients of $\hat{u}$ are compressed from 14 bits ($r_{\hat{u}}=q$) to 8 bits ($r_{\hat{u}}=256$) when the conventional compression rate ($r_{v'}=8$) for the coefficients of $v'$ is applied.
This can reduce the ciphertext size by 35.3 \%.
Furthermore, the compression for the coefficients of $ \hat{u} $ and additional compression for the coefficients of $v'$, which is Option 2 and Option 3 used with (4,256), can reduce the ciphertext size by 41.2 \% to improve the bandwidth efficiency while achieving the target DFR.
If the goal is to reduce the ciphertext size by 35.3 \%, (4,512) is much better than (8,256) because (4,512) shows a remarkably lower DFR than (8,256).
Especially, because Option 1 used with (8,512) and (4,512), Option 2 used with (8,512), (8,256), and (4,512), and Option 3 used with (8,512), (8,256), (4,512), and (4,256) overachieve the target DFR, this DFR margin can also be exploited to improve security of NewHope by increasing the noise parameter $k$.

\begin{table}[h]
\centering

\caption{DFR and ciphertext size of NewHope when Option 4 is used with various compression rate when $n=512$.}
\begin{tabular}{|c|l|l|c|}
\hline
\multicolumn{1}{|c|}{($r_{v'}$, $r_{\hat{u}}$)}   & \multicolumn{1}{c|}{ECC Option} & \multicolumn{1}{c|}{DFR} & \begin{tabular}[c]{@{}c@{}}Ciphertext size\\ (reduction rate)\end{tabular} \\ \hline
\multirow{2}{*}{(8,512)}  & Only ATE (NewHope)          & $\approx 2^{-33}$                & \multirow{2}{*}{\begin{tabular}[c]{@{}l@{}}\multicolumn{1}{c}{768 bytes}\\ (29.4 \%)\end{tabular}}    \\ \cline{2-3}
                                                  & Option 4                    & $\approx 2^{-1101}$              &                                                                                              \\ \hline
\multirow{2}{*}{(8,256)}  & Only ATE (NewHope)          & $\approx 2^{-7}$                 & \multirow{2}{*}{\begin{tabular}[c]{@{}l@{}}\multicolumn{1}{c}{704 bytes}\\  (35.3 \%)\end{tabular}}    \\ \cline{2-3}
                                                  & Option 4                    & $\approx 2^{-295}$               &                                                                                              \\ \hline
\multirow{2}{*}{(8,128)}  & Only ATE (NewHope)          & $\approx 2^{-1}$                  & \multirow{2}{*}{\begin{tabular}[c]{@{}l@{}}\multicolumn{1}{c}{640 bytes}\\  (41.2 \%)\end{tabular}}    \\ \cline{2-3}
                                                  & Option 4                    & $\approx 2^{-21}$                  &                                                                                              \\ \hline
\multirow{2}{*}{(4,1024)} & Only ATE (NewHope)          & $\approx 2^{-43}$                & \multirow{2}{*}{\begin{tabular}[c]{@{}l@{}}\multicolumn{1}{c}{768 bytes}\\  (29.4 \%)\end{tabular}}    \\ \cline{2-3}
                                                  & Option 4                    & $\approx 2^{-2842}$              &                                                                                              \\ \hline
\multirow{2}{*}{(4,512)}  & Only ATE (NewHope)          & $\approx 2^{-15}$                & \multirow{2}{*}{\begin{tabular}[c]{@{}l@{}}\multicolumn{1}{c}{704 bytes}\\  (35.3 \%)\end{tabular}} \\ \cline{2-3}
                                                  & Option 4                    & $\approx 2^{-539}$               &                                                                                              \\ \hline
\multirow{2}{*}{(4,256)}  & Only ATE (NewHope)          & $\approx 2^{-8}$                  & \multirow{2}{*}{\begin{tabular}[c]{@{}l@{}}\multicolumn{1}{c}{640 bytes}\\ (41.2 \%)\end{tabular}}    \\ \cline{2-3}
                                                  & Option 4                    & $\approx 2^{-138}$                                                                                                               & \\ \hline
\end{tabular}
\label{Bandwidth efficiency improvement using ECC 512}
\end{table}

Table \ref{Bandwidth efficiency improvement using ECC 512} shows the DFR and ciphertext reduction of NewHope with $n=512$ when Options 4 is used with various compression rates given in Table \ref{Compression rates table}.
(4,512) is better than the (8,256) because (4,512) and (8,256) reduce the ciphertext size equally, but (4,512) shows a significantly lower DFR than (8,256).
Because Option 4 used with (8,512), (8,256), (4,1024), and (4,512) overachieves the target DFR, this DFR margin can be exploited to improve the security of NewHope by increasing the noise parameter $k$.
We will call the options that overachieve the target DFR the excellent ECC candidates to be used in the next section.

%Previous Section에서 제시한 다양한 compression rate의 후보들로 bandwidth efficiency를 향상시킬 수 있지만, ATE만으로는 PQC에 적합한 DFR을 달성하지 못한다. 이점을 극복하기 위해서 BCH code와 ATE를 연접하여 PQC에 적합한 DFR을 달성할 수 있다. $\hat{u}$에 대해서만 14 bits를 8 bits로 압축하여 기존의 ciphertext size에 비해 35.3 \% 를 감소시키면서도 BCH code가 적용된 모든 option들에 대해서 $2^{-151}$ 이하의 DFR을 달성할 수 있다. (2176 bytes \rightarrow 1408 bytes). Bandwidth efficiency를 더 개선시키기 위해서는 $v'$에 대해서도 추가적인 압축을 수행하여 기존의 ciphertext size에 비해 41.2 \%로 감소시키면서도 Option 2, 3을 통해 $2^{-168}$ 이하의 DFR을 달성할 수 있다. 그리고 다른 결과들 중에는 지나치게 낮은 DFR을 달성하는 것들도 있다. For example, Option 3 when ($r_{v'}=8$, $r_{\hat{u}}=512$ and Option 3 when ($r_{v'}=4$, $r_{\hat{u}}=512$ and so on. 이러한 것들의 과도한 DFR 성능을 활용하여 security를 향상시킬 수 있다.

\subsection{Security Analysis for The Excellent ECC Candidates}

The excellent ECC candidates obtained in section 4.2 show excessive DFR performance owing to the properly chosen compression rate and ECC.
Such overachieved DFR can be exploited to improve the security level of NewHope.

\begin{table}[h]
\centering
\caption{Improved security level, DFR, and ciphertext size for excellent ECC candidates for NewHope with $n=1024$.}

\begin{tabular}{|c|c|c|c|c|c|c|}
\hline
\multirow{2}{*}{($r_{v'}$,$r_{\hat{u}}$)} & \multirow{2}{*}{ECC Option} & \multirow{2}{*}{\begin{tabular}[c]{@{}c@{}}Ciphertext size\\ (reduction rate)\end{tabular}} & \multirow{2}{*}{$k$} & \multicolumn{2}{c|}{Classical/Quantum}                                                                                  & \multirow{2}{*}{DFR} \\ \cline{5-6}
                                         &                             &                                                                                             &                      & Primal {[}bits{]}   & Dual {[}bits{]} &                      \\ \hline
(8,$q$)                                  & NewHope                     & 2176 bytes                                                                                  & 8                    & 259/235                                                     & 257/233                                                   & $\approx 2^{-474}$   \\ \hline
\multirow{3}{*}{(8,512)}                 & Option 1                    & \multirow{3}{*}{\begin{tabular}[c]{@{}c@{}}1536 bytes\\ (29.4 \%)\end{tabular}}             & 23                   & 296/268                                                     & 294/267                                                   & $\approx 2^{-147}$   \\ \cline{2-2} \cline{4-7} 
                                         & Option 2                    &                                                                                             & 33                   & 310/281                                                     & 309/280                                                   & $\approx 2^{-144}$   \\ \cline{2-2} \cline{4-7} 
                                         & Option 3                    &                                                                                             & 36                   & 314/285                                                     & 312/283                                                   & $\approx 2^{-140}$   \\ \hline
\multirow{3}{*}{(8,256)}                 & Option 1                    & \multirow{3}{*}{\begin{tabular}[c]{@{}c@{}}1408 bytes\\ (35.3 \%)\end{tabular}}             & 8                    & 259/235                                                     & 257/233                                                   & $\approx 2^{-151}$   \\ \cline{2-2} \cline{4-7} 
                                         & Option 2                    &                                                                                             & 13                   & 275/249                                                     & 274/248                                                   & $\approx 2^{-145}$   \\ \cline{2-2} \cline{4-7} 
                                         & Option 3                    &                                                                                             & 14                   & 278/252                                                     & 276/250                                                   & $\approx 2^{-158}$   \\ \hline
\multirow{3}{*}{(4,512)}                 & Option 1                    & \multirow{3}{*}{\begin{tabular}[c]{@{}c@{}}1408 bytes\\ (35.3 \%)\end{tabular}}             & 15                   & 280/254                                                     & 279/253                                                   & $\approx 2^{-151}$   \\ \cline{2-2} \cline{4-7} 
                                         & Option 2                    &                                                                                             & 25                   & 299/271                                                     & 298/270                                                   & $\approx 2^{-140}$   \\ \cline{2-2} \cline{4-7} 
                                         & Option 3                    &                                                                                             & 27                   & 302/274                                                     & 301/273                                                   & $\approx 2^{-146}$   \\ \hline
\multirow{2}{*}{(4,256)}                 & Option 2                    & \multirow{2}{*}{\begin{tabular}[c]{@{}c@{}}1280 bytes\\ (41.2 \%)\end{tabular}}             & 9                    & 262/238                                                     & 261/237                                                   & $\approx 2^{-143}$   \\ \cline{2-2} \cline{4-7} 
                                         & Option 3                    &                                                                                             & 10                   & 266/241                                                     & 265/240                                                   & $\approx 2^{-147}$   \\ \hline
\end{tabular}
\label{Security improvement using ECC}
\end{table}

Tables \ref{Security improvement using ECC} shows the DFR, ciphertext size, and security level, which is estimated at the cost of primal attack and dual attack of NewHope with $n=1024$.
Compared with current NewHope with $n=1024$, Option 3 used with (8,512) can improve the security level by 21.5 \% and reduce the ciphertext size by 29.4\% while achieving the target DFR $2^{-140}$.
If we focus on improving the bandwidth efficiency, Option 3 used with (4,256) improves the security level by 3 \% as well as reduces the ciphertext size by 41.5 \% while achieving the target DFR.
Option 3 used with (4,512) and Option 3 used with (8,256) improve the bandwidth efficiency equally, but  the former one is better because it improves the security level by 9.4 \% more than latter one.

\begin{table}[h]
\centering
\caption{Improved security level, DFR, and ciphertext size for excellent ECC candidates for NewHope with $n=512$.}

\begin{tabular}{|c|c|c|c|c|c|c|}
\hline
\multirow{2}{*}{($r_{v'}$,$r_{\hat{u}}$)} & \multirow{2}{*}{ECC Option} & \multirow{2}{*}{\begin{tabular}[c]{@{}c@{}}Ciphertext size\\ (reduction rate)\end{tabular}} & \multirow{2}{*}{$k$} & \multicolumn{2}{c|}{Classical/Quantum}                                                                                  & \multirow{2}{*}{DFR} \\ \cline{5-6}
                                         &                             &                                                                                             &                      & Primal {[}bits{]}   & Dual {[}bits{]} &                      \\ \hline
(8,$q$)                                  & NewHope                     & 1088 bytes                                                                                  & 8                    & 112/101                                                     & 112/101                                                   & $\approx 2^{-431}$   \\ \hline
(8,512)                                  & Option 4                    & \begin{tabular}[c]{@{}c@{}}768 bytes\\ (29.4 \% reduction)\end{tabular}                     & 33                   & 137/124                                                     & 137/124                                                   & $\approx 2^{-140}$   \\ \hline
(8,256)                                  & Option 4                    & \begin{tabular}[c]{@{}c@{}}704 bytes\\ (35.3 \% reduction)\end{tabular}                     & 13                   & 120/109                                                     & 119/108                                                   & $\approx 2^{-151}$   \\ \hline
(4,512)                                  & Option 4                    & \begin{tabular}[c]{@{}c@{}}704 bytes\\ (35.3 \% reduction)\end{tabular}                     & 22                   & 129/117                                                     & 129/117                                                   & $\approx 2^{-145}$   \\ \hline
\end{tabular}
\label{Security improvement using ECC 512}
\end{table}

Tables \ref{Security improvement using ECC 512} shows the DFR, ciphertext size, and security level of NewHope with $n=512$.
Compared with the current NewHope with $n=512$, Option 4 used with (8,512) can improve the security level by 22.8 \% and reduce the ciphertext size by 29.4 \% while achieving the target DFR.
Additionally, Option 4 used with (4,512) can improve the security level by 15.8 \% and reduce the ciphertext size by 35.3 \% while achieving the target DFR.
The complexity of the decoding algorithm (BM algorithm) of BCH codes increases linearly with the number of correctable errors $C_t$.
Thus, if the decoding complexity of BCH code is important, it is better to choose Option 1 used with (8,512) or Option 1 used with (4,512).

%NewHope는 Ring-LWE problem에 기반한 cryptosystem이다. The original worst-case to average-case reduction to Ring-LWE states hardness for rounded Gaussian distribution. NewHope는 rounded Gaussian distribution 대신 centered binomial distribution을 사용하기 때문에 security 감소는 불가피하다. 그러나 Section 2.3에서 noise parameter $k$가 증가함에 따라 centered binomial distribution과 rounded Gaussian distribution의 SD가 작아짐을 확인하였다. 따라서 NewHope에 concatenated coding schemes를 적용함으로써 충분히 큰 noise parameter $k$에 대해서도 the target DFR을 달성할 수 있고 이로 인해 NewHope의 hardness도 개선시킬 수 있다.

% BCH code의 강력한 오류 정정 능력으로 인해 과도하게 DFR 성능을 달성한 후보들이 있다. 과도하게 달성된 DFR을 이용하여 security level을 증가시킬 수 있다. 이것은 강력한 ECC를 적용함으로써 얻을 수 있는 가장 큰 장점이다. 표는 DFR을 초과 달성한 후보들에 대해서 noise variance를 증가시키면서 PQC에 적합한 DFR을 달성하는 결과를 보여준다. 표의 security level은 primal attack과 dual attack 중에서 낮은 cost를 갖는 것에 대한 결과만을 나타내었다. NewHope에 BCH code를 적용하여 security level을 최대 21.5 \% 증가시킬 수 있으면 동시에 ciphertext는 29,4 \% 감소시킬 수 있다. NewHope의 성능 개선을 bandwidth efficiency에 초점을 맞춘다면, ciphertext를 35.3 \% 감소시키면서 security level를 17.1 \% 증가시키실 수 있다.

\subsection{Closeness of Centered Binomial Distribution and the Corresponding Rounded Gaussian Distribution for Various $k$}

The properties of rounded Gaussian distribution $\xi$ are key to the worst-case to average-case reduction for Ring-LWE.
However, since a very high-precision sampling is required for the rounded Gaussian distribution, NewHope uses the centered binomial distribution $\psi_k$ for practical sampling without having rigorous security proof.
It is generally accepted that as the centered binomial distribution and the rounded Gaussian distribution are closer to each other, NewHope is regarded as more secure. 
%Centered binomial distribution과 rounded Gaussian distribution의 statistical distance (SD)가 작을 수록 NewHope의 security가 개선된다고 할 수 있다. 
The closeness of two distribution can be measured through many methods.
Among them, R{\'e}nyi divergence is a well-known method, which is parameterized by a real $a>1$ and defined for two distributions $P$ and $Q$ as follows.

\begin{equation}
R_a(P||Q)=\Bigg( \sum_{x \in \sup(P)} \frac{P(x)^a}{Q(x)^{a-1}} \Bigg)^{\frac{1}{a-1}}
\end{equation}
where $sup(P)$ represents the support of $P$ and $Q(x) \neq 0$ for $x \in sup(P)$.

We define $\xi_k$ to be the rounded Gaussian distribution with the variance $\sigma^2=k/2$, which is the distribution of $\lfloor \sqrt{k/2} \cdot x \rceil$ where $x$ follows the standard normal distribution.

\begin{figure}[h]
\centering
\includegraphics[width=\textwidth]{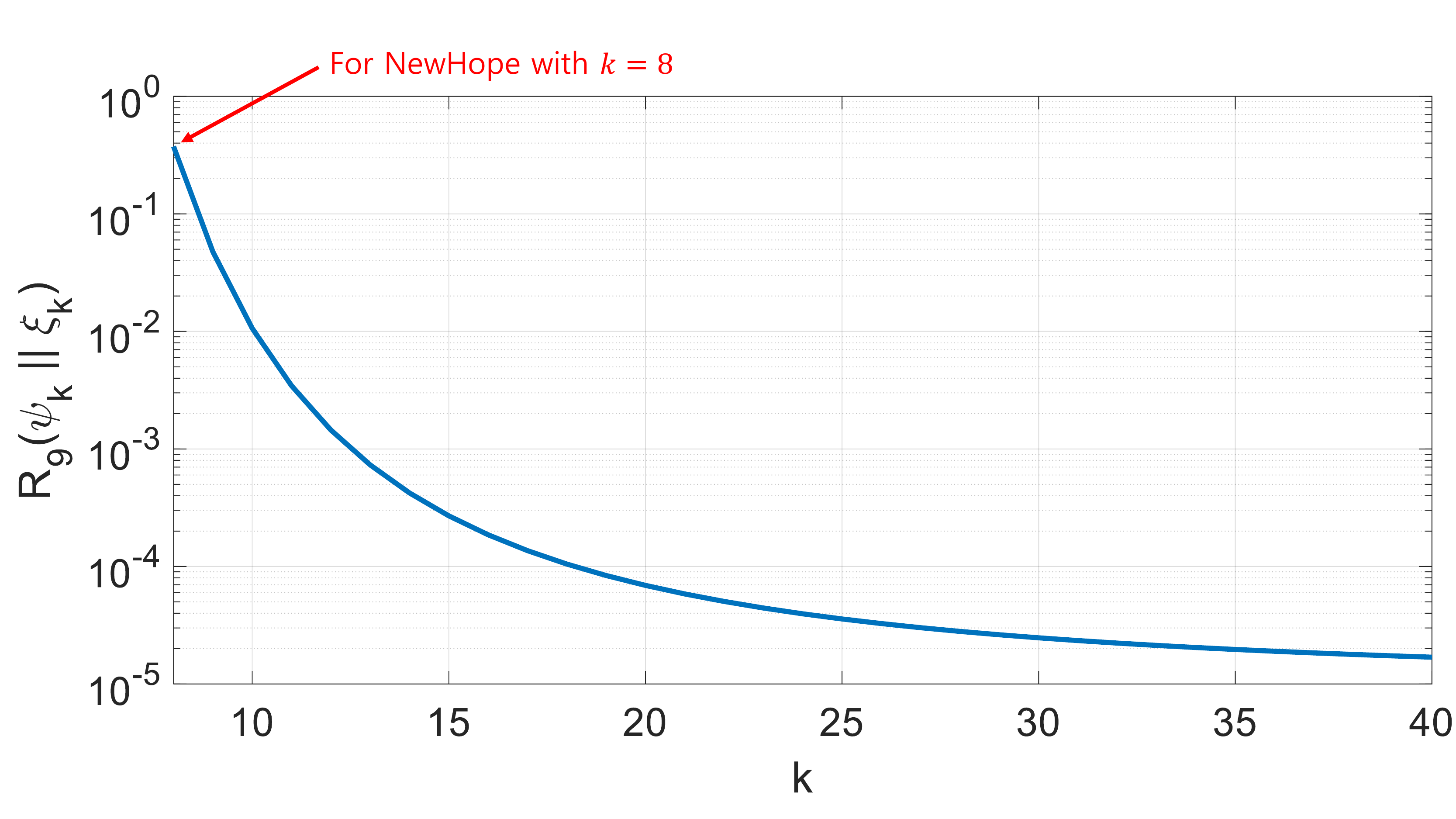}
\caption{R{\'e}nyi divergence of centered binomial distribution $\psi_k$ and rounded Gaussian distribution $\xi_k$ with the same variance $k/2$ according to $k$ ($a=9$).} 
\label{Renyi}
\end{figure}

Fig. \ref{Renyi} shows that the R{\'e}nyi divergence ($a=9$ is used as in \cite{NewHope NIST}) of the centered binomial distribution $\psi_k$ and rounded Gaussian distribution $\xi_k$ with the same variance $k/2$ decreases as $k$ increases.
Therefore, an increase in noise parameter $k$ can quantitatively and qualitatively improve the security of NewHope although the time complexity increases due to the complexity increase of $\sum_{i=0}^{k-1}(b_i-b_i')$.

%Excellent candidates 
By applying concatenated coding schemes to NewHope, the target DFR can be achieved even by using a large noise parameter $ k $, which therefore improves the security of NewHope due to improved closeness of centered binomial distribution and the corresponding rounded Gaussian distribution.

\section{Conclusions and Future Works}
In this paper, it is shown that the security and bandwidth efficiency of NewHope can be substantially improved by doing exact noise analysis and adopting proper concatenated error-correction schemes of ATE and BCH code.
In detail, the DFR of NewHope is recalculated as $2^{-474}$ and $2^{-431}$ for $n=1024$ and $n=512$ through exact noise analysis, and hence it is shown that the DFR of NewHope has been too conservatively calculated.
Since the recalculated DFR is much lower than the required $2^{-140}$, this DFR margin is exploited to improve the security level of NewHope by 8.5\% with little increase in time/space complexity.
Also, this margin is utilized to improve the bandwidth efficiency by 5.9 \% with keeping the time/space complexity of NewHope.
Furthermore, by allowing a slight increase in time/space complexity, the bandwidth efficiency can be improved by 23.5 \% using an additional compression on $\hat{u}$ while achieving the required DFR.

The ATE of NewHope is simple and robust to the side channel attack, but its error-correction capability is relatively weak compared with other ECCs.
Therefore, we propose various concatenated coding schemes of ATE and BCH code to combine the advantages of these two ECCs.
From an information-theoretic viewpoint, ATE and total noise channel can be regarded as a super channel, and the security and bandwidth efficiency improvement can be analyzed based on super channel analysis.
By applying selected concatenated coding schemes to NewHope, the security level can be improved up to 21.5 \% and the bandwidth efficiency can be improved up to 41.2 \% by reducing ciphertext size, compared with the current NewHope with $n=1024$.
Likewise, the security level can be improved up to 22.8 \% and the bandwidth efficiency can be improved up to 35.3 \% by reducing ciphertext size, compared with the current NewHope with $n=512$.
Furthermore, by applying concatenated coding schemes to NewHope, the security of NewHope can be enhanced by improved closeness of centered binomial distribution and the corresponding rounded Gaussian distribution 
the target DFR can be achieved even by using a large noise parameter $ k $, which therefore improves 
However, BCH codes are less robust to the side channel attack than ATE and have a high implementation complexity.
For this reason, several improvements are required for BCH codes to be practically used in NewHope or other lattice-based cryptosystems.
Moreover, optimizing ECC parameters will improve the security and bandwidth efficiency of NewHope.

% 본 논문에서는 NewHope에 대해서 PQC로써 가장 중요한 덕목인 security, cost, and performance를 정확한 noise analysis으로 개선시킨다. 기존에 계산되었던 DFR에 대한 부족한 부분을 보완하여 재계산하였으며, 이를 통해 과도한 DFR 성능을 활용하여 복잡도 증가없이 security를 8.5 \% 개선시킨다. 뿐만 아니라 약간의 복잡도를 증가시키만 ciphertext size를 최대 5.9 \% 까지 감소시켜 bandwidth efficiency를 개선시킨다.
% 본 논문에서는 기존에 Ring LWE기반의 암호의 파라마터로써 정의된 dimension, modulus, and noise vairance 에 추가로 bandwidth efficiency와 관계있는 compression rate에 대해서 정의한다. 다양한 compression rate를 활용하여 보다 확장된 개념으로 NewHope의 security와 bandwidth efficiency를 개선시킨다. NewHope는 구현이 용이하고 side channel attack에 강인한 ATE를 ECC로 활용하였다. 그러나 낮은 오류 정정 능력을 갖는 단점이 있고, 이를 강력한 오류 정정 능력을 갖는 BCH code와 연접을 통해 NewHope의 security와 bandwidth efficiency를 개선시킨다. NewHope에 새롭게 정의한 compression rate와 더불어 ATE와 BCH code의 연접을 통해 기존 NewHope에 비해 security를 최대 21.5 \% 만큼 개선하였으며, bandwidth efficiency를 최대 35.3 \% 까지 개선시켰다.
% 그러나 BCH code는 ATE만큼 side channel attack에 강인하지 못하기 때문에 구현 관점에서는 추후 개선할 여지가 있다. 또한 BCH code의 parameter를 NewHope에 최적화하여 추후 성능을 개선할 여지가 있다.

%
% ---- Bibliography ----
%
% BibTeX users should specify bibliography style 'splncs04'.
% References will then be sorted and formatted in the correct style.
%
% \bibliographystyle{splncs04}
% \bibliography{mybibliography}

\end{document}